\documentclass[prl, twocolumn]{revtex4-1}

\usepackage{graphicx,graphics,color,epsfig}
\usepackage{amsmath}
\usepackage{amssymb}
\usepackage{amsfonts}
\usepackage{amsfonts}
\usepackage{epstopdf}
\usepackage{bm}
\usepackage{times,xspace}
\usepackage[colorlinks,citecolor=blue,linkcolor=red]{hyperref}
\usepackage{color}

\begin{document} 
 
\title{Topological Landau-Zener Bloch Oscillations in Photonic Floquet Lieb Lattices}

\author{Yang Long}
\author{Jie Ren}
 \email{Xonics@tongji.edu.cn}
\affiliation{%
Center for Phononics and Thermal Energy Science, China-EU Joint Center for Nanophononics, Shanghai Key Laboratory of Special Artificial Microstructure Materials and Technology,
School of Physics Sciences and Engineering, Tongji University, Shanghai 200092, China
}%

\date{\today}


\maketitle

{\bf The Lieb Lattice exhibits intriguing properties that are of general interest in both the fundamental physics and practical applications. Here, we investigate the topological Landau-Zener Bloch oscillation in a photonic Floquet Lieb lattice, where the dimerized helical waveguides is constructed to realize the synthetic spin-orbital interaction through the Floquet mechanism, rendering us to study the impacts of topological transition from trivial gaps to non-trivial ones. The compact localized states of flat bands supported by the local symmetry of Lieb lattice will be associated with other bands by topological invariants, Chern number, and involved into Landau-Zener transition during Bloch oscillation. Importantly, the non-trivial geometrical phases after topological transitions will be taken into account for constructive and destructive interferences of wave functions. The numerical calculations of continuum photonic medium demonstrate reasonable agreements with theoretical tight-binding model. Our results provide an ongoing effort to realize designed quantum materials with tailored properties.}

The study of topological states of matter\cite{Moore2010The} have been of huge interest since the theoretical prediction and experimental observation of the quantum spin Hall effect\cite{Hasan2010Colloquium}. Besides the novel properties of quantum Hall and quantum spin Hall effect, other topological states have been also studied widely, e.g, topological superconductors\cite{qi2011topological}, magnetic monopoles\cite{Qi2009Inducing} and Weyl points in topological semi-metals\cite{Xu2015TOPOLOGICAL}. The topological states not only have been studied in electronic crystal, but also have been extended and explored in the photonics\cite{Rechtsman2013Photonic} and acoustic field\cite{yang2015topological}. Photonic topological insulators (PTIs)\cite{Lu2014Topological} are photonic devices with many novel topological properties, e.g, spin-dependent unidirectional edge states and topological non-trivial gapped photonic band structures, analogous to single-particle electronic band structures of topological insulators. Recently, the PTIs have been identified in many setups, including magneto-optic photonic crystals\cite{Rechtsman2013Photonic}, ring resonator lattices\cite{Liang2012Optical}, electronics circuits\cite{ningyuan2015time}, lattice symmetry operations\cite{wu2015scheme}, hyperbolic chiral metamaterials\cite{gao2015topological}. Specially, in the potentially useful optical frequency regime, the Floquet topological insulator (FTI)\cite{rechtsman2013floquet} has been proposed theoretically and demonstrated experimentally through arrays of helical optical waveguides. The physical mechanism behind FTI can be described by the quantum systems with time-periodic Hamiltonians\cite{kitagawa2010topological}, in which the periodic driving can induce topological non-trivial states without external magnetic field or spin-orbit effects in conventional static Hamiltonian\cite{struck2012tunable}. Besides the expected properties of topological matters, the additional interactions between Floquet mechanism and lattice systems would lead to interesting new phenomena, i.e, anomalous topological phases\cite{leykam2016anomalous}.

Meanwhile, flat bands (FBs), a class of completely dispersionless bands in the spectrum, have attracted lots of research attention recently.  FBs have been explored theoretically and experimentally in, e.g, two-dimensional optical Lieb waveguide lattices\cite{guzman2014experimental,vicencio2015observation,mukherjee2015observation,diebel2016conical},  flat-band Hubbard models\cite{mielke1999ferromagnetism}, and fractional Chern insulators\cite{neupert2011fractional,liu2013flat}. The existence of FBs is ensured by the local symmetries in specific translationally invariant lattices, and their behavior are similar to ``dark'' states that cause the coherent trapping in quantum optics. Indeed, FBs rely on compact localized states due to destructive wave interference induced by specific lattice geometries. Compact localized states\cite{flach2014detangling} can hold lots of novel properties, e.g, infinite effective mass in electronic bands\cite{jacqmin2014direct}, localized states in photonics networks\cite{vicencio2015observation,mukherjee2015observation}, and unusual ferromagnetic ground states\cite{mielke1999ferromagnetism}. Therefore, it is interesting to investigate the interplay between the periodic-driving Floquent mechanism and compact localized states of flat bands, as well as the induced nontrivial properties. 

In this work, we demonstrate optical Landau-Zener Bloch oscillation in a photonic Floquet Lieb lattice, which manifests the combined physics of topological insulators, Bloch oscillations, Landau-Zener tunneling\cite{zener1932non, khomeriki2016landau}, and compact localized states. 
Our design is based on  Lieb lattices of helical waveguides, with adjusting the nearest neighbor coupling strength to realize topological non-trivial band gaps and one perturbed flat band. The periodic modulation arisen from the helical twisting of the photonic waveguides can generate artificial gauge field and synthetic spin-orbital interaction of light to achieve non-zero magnetic flux in sublattices through the Floquet mechanism~\cite{rechtsman2013floquet}. Two types of band gaps, topological trivial and non-trivial ones, are achieved to investigate the topological effect to flat bands. Due to non-trivial topological transition, the compact localized states of flat bands will be associated with other bands by topological invariants, i.e, Chern number, and thus these states will involve in Landau-Zener transitions. Different from the conventional Bloch oscillation determined by the total phases accumulated during adiabatic and non-adiabatic process, namely, St\"uckelberg phases\cite{kling2010atomic}, the Floquet topological Lieb insulator demonstrates that the non-trivial geometrical phase will make an additional interference with the these phases.

\begin{figure}
\centering
\includegraphics[width=\linewidth]{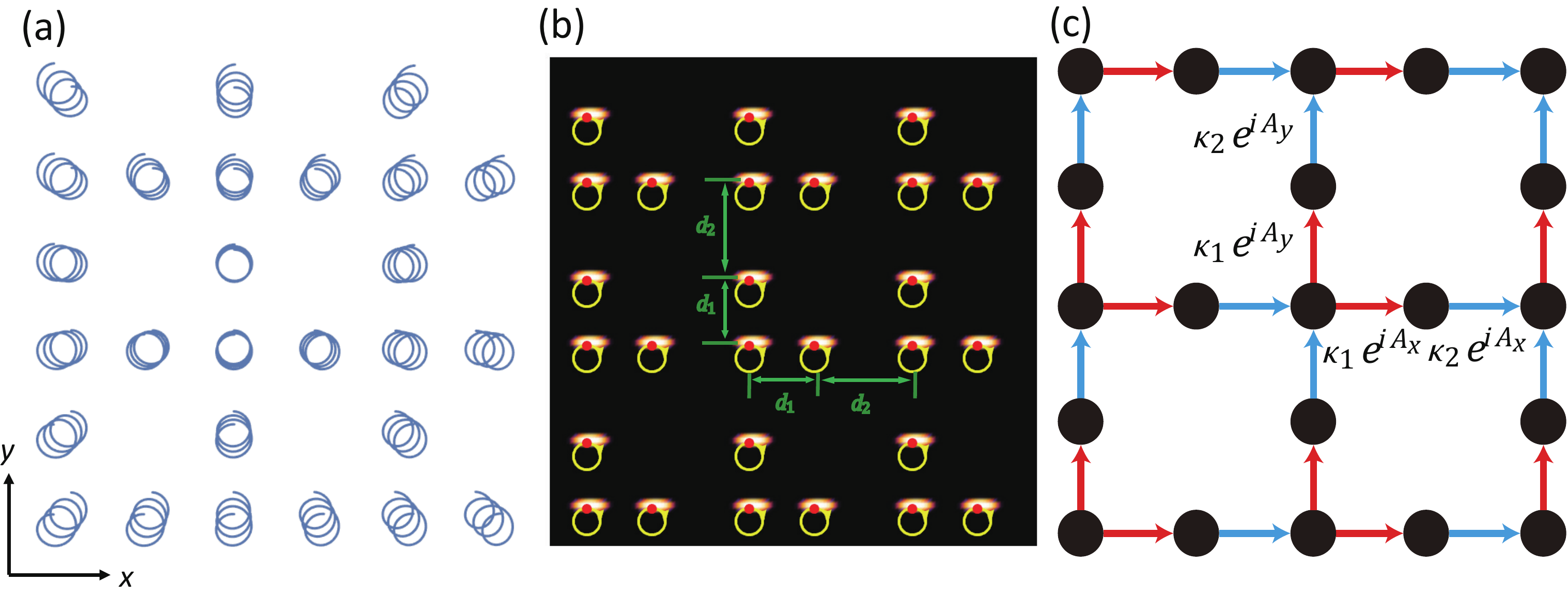}
\caption{ {\bf Topological Lieb lattice of helical waveguides.} (a) Scheme of helical waveguides. Their rotation axis is along the $z$ direction, with specially appointed separation, uniform helix radius $R$ and the period $Z_0$. (b) Cross section of the Lieb lattice of helical waveguides, which have circular trajectories with the variation of $z$. (c) Equivalent tight-binding model for describing the dynamic properties of Floquet mechanism. The coupling between helical waveguides can be considered as the nearest-neighbor interaction under an effective gauge field $\bm{A}(z)$.}
\label{fig:floquetlieblattice}
\end{figure}

\vskip0.25cm\noindent
{\bf Results}\\
{\bf Floquet Lieb lattices with helical photonic waveguides.}
Paraxial light propagation in photonic lattices can be described by a paraxial field $\psi(x,y,z)$ governed by the Schr$\ddot{o}$dinger-type equation:
\begin{equation}
i\frac{\partial \psi}{\partial z} = -\frac{1}{2n_0k_0} \nabla^2\psi(x,y,z) - k_0\Delta n(x,y,z)\psi(x,y,z)
\label{eq:schrodingerequation}
\end{equation}
where $\nabla_{\bot}^2 = \partial_x^2 + \partial_y^2$, $k_0 = 2\pi/\lambda_0$ and $\lambda_0$ is the wavelength in vaccum. After transforming the coordinate into a reference frame where the waveguides are invariant in the $z$ direction, namely: $x' = x + R \sin(\Omega z)$, $y'=y+R \sin(\Omega z - \phi)$ and $z'=z$, where $R$ is helix radius, $Z_0 = 2\pi/\Omega$ is period and $\phi$ is the delay phase ,  we can induce an effective gauge field $\bm{A}(z)$ for photonics crystal.

Our system of current interest is the topological Lieb lattice shown in Fig.\ref{fig:floquetlieblattice}(a)(b). The $xOy$ position of each helix in unit cell of Lieb Lattice is shifted along $x$ or $y$ axis, respectively. The dimerized form-shifts produce a topological trivial gap, and server to build a perturbed flat band when coupling with Floquet mechanism. Similar schemes can be implemented in other lattice geometries, such as a Kagome lattice. Without loss of generality, we focus on the Lieb lattice here.

Tight-binding models\cite{morandotti1999experimental,diebel2016conical,vicencio2015observation,peschel1998optical} are known to describe qualitatively well in a realistic photonic lattice that obeys paraxial wave equation, such as femtosecond laser-written waveguides in fused silica. Thus we can focus on the tight-binding model to investigate the intrinsic mechanism. Our model system consists of a two-dimensional photonic lattice as shown in Fig. (\ref{fig:floquetlieblattice})c. The beam light propagates along $z$ and the period of the helical waveguides is sufficiently small that guided mode is adiabatically drawn along with a waveguide as it curves. Thus our tight-binding model can be considered to be modulated by a external effective gauge field $\bm{A}(z)$. Each unit cell in the Lieb lattice has three sites with the different coordinates, and under periodic modulations, the Hamiltonian $H(z)$ reads:
\begin{equation}
\begin{split}
H(z) = \sum_{\langle n,m \rangle} (\kappa_{1} e^{iA_y(z)} b^{\dagger}_{n,m} a_{n,m} + \kappa_{2} e^{-iA_y(z)} b^{\dagger}_{n,m-1} a_{n,m}  \\ + \kappa_{1} e^{iA_x(z)} c^{\dagger}_{n,m} a_{n,m} + \kappa_{2} e^{-iA_x(z)} c^{\dagger}_{n-1,m} a_{n,m}) + h.c.
\label{eq:liebHamiltonian}
\end{split}
\end{equation}
where $a_{m,n}^{\dagger}(a_{m,n})$,$b_{m,n}^{\dagger}(b_{m,n})$, and $c^{\dagger}(c_{m,n})$ are the creation(annihilation) operators of sites $a$,$b$, and $c$ in lattice unit $(n,m)$, respectively. $\kappa_{1}$,$\kappa_{2}$ are hopping energy, in our model, $\kappa_{1} \ne \kappa_{2}$. $A_x(z)$ and $A_y(z)$ are the components of periodic-driven gauge field $\bm{A}(z)$ on axis $x$ and $y$, respectively. Following the waveguide helicity depicted after Eq.~\ref{eq:schrodingerequation}, the periodic-driven gauge field $\bm{A}(z)$ has the following modulation form:
\begin{equation}
A_x = A_0 \sin(\Omega z),   A_y = A_0 \sin(\Omega z - \phi)
\label{eq:gaugefield}
\end{equation}
where $A_0=k_0R\Omega$ is the synthetic gauge field strength. 

For Floquet system, the periodic driven Schr\"odinger equation can be solved by a Fourier expansion, $H(z) = H_0 + H_1 e^{i \Omega z} + H_{-1} e^{-i \Omega z} $. For simplicity, consider the coefficients of high order $n$th ($|n| \geq 2 $) Fourier components are small enough to ignore. Following the same spirit of Ref.~\cite{cayssol2013floquet}, for small modulation strength $J_0(A_1) \ll J_1(A_0)$, where $J_0$ and $J_1$ are bessel functions with order 0 and 1, defining $H_n = \frac{1}{Z_0}\int_{0}^{Z_0} e^{i n \Omega z} H(\bm{k},\bm{A}(z)) dz$ as the $n$-th Fourier harmonic of the periodic-driven Hamiltonian $H(\bm{k},\bm{A}(z))$, the effective Hamiltonian of this periodic-modulated system can be written as $ H_{eff} = H_0 + \frac{1}{\Omega}[H_1,H_{-1}]$, such that: 
\begin{widetext}
\begin{equation}
H_{eff} = \Delta_0 \begin{pmatrix}
0 & \kappa_1 e^{i k_x} + \kappa_2 e^{-i k_x} & i\tau (\kappa_1 e^{i k_x} - \kappa_2 e^{- i k_x})(\kappa_1 e^{i k_y} - \kappa_2 e^{-i k_y}) \\
\kappa_1 e^{-i k_x} + \kappa_2 e^{i k_x}  & 0 & \kappa_1 e^{i k_y} + \kappa_2 e^{-i k_y} \\
- i\tau (\kappa_1 e^{-i k_x} - \kappa_2 e^{ i k_x})(\kappa_1 e^{-i k_y} - \kappa_2 e^{i k_y}) & \kappa_1 e^{-i k_y} + \kappa_2 e^{i k_y} & 0
\end{pmatrix}
\label{eq:effectivehamltonian}
\end{equation}
\end{widetext}
where $\Delta_0 = J_0(A_0)$, $\Delta_1 = -\frac{J_1(A_0)^2 sin(\phi)}{2 \omega}$, and $\tau=\frac{\Delta_1}{\Delta_0}$  (More details in Supplement). 
Since $H_{eff}$ is effective Hamiltonian for periodic-driven system, its eigenvalues have the form $e^{i \beta(\bm{k})}$ where $\beta(\bm{k})$ is the quasi-energy spectrum. $H_{eff}$ reveals all the interactions between compact localized state of Lieb lattice system under periodic modulations.

\begin{figure}
\centering
\includegraphics[width=\linewidth]{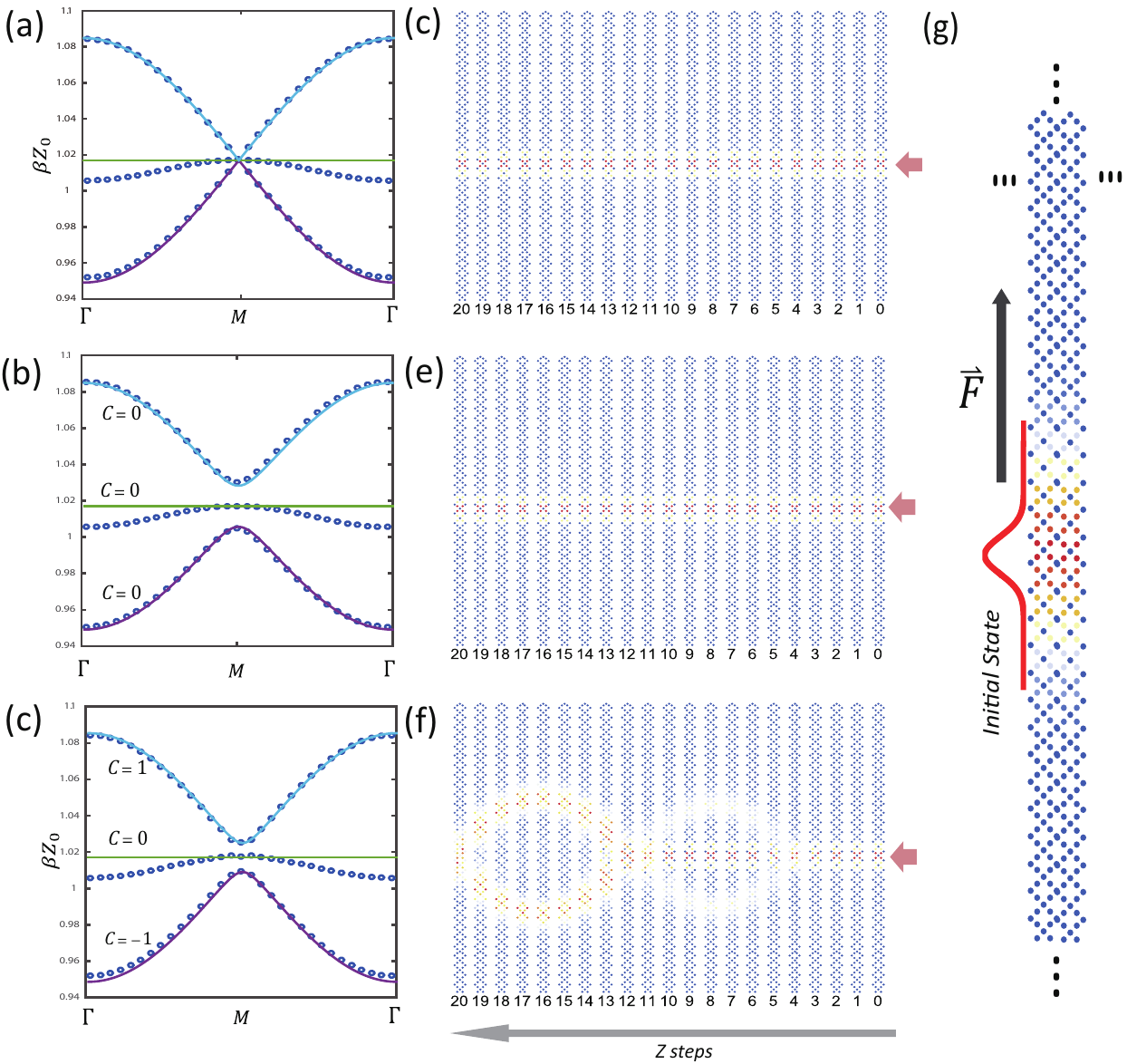}
\caption{{\bf Quasi-Energy band structures and Landau-Zener Bloch oscillations.} (a, d) Standard Lieb lattice of waveguides ($\kappa_1=\kappa_2$, $A_0=0$). (b, e) Dimerized Lieb lattice of waveguides($\kappa_1\ne \kappa_2$, $A_0=0$). (c, f) Floquet Topological Lieb lattice of helical waveguides($\kappa_1= \kappa_2$, $A_0\ne0$).  In band structure, blue circle points are obtained from the continuum model and solid curves from the tight-binding model with parameters fitted from numerical results. 
Landau-Zener Bloch oscillations are numerically simulated with tight-binding model by exciting initially a Gaussian flat band state as shown in (g). Initial flat band state stays the same position and no oscillation occur when $A_0=0$. For the case with nonzero effective guage fields, the initial ``dark'' flat state starts to evolve. The Bloch oscillation of period $5Z_0$ is displayed, and the evolution of the norm density of sites in sublattice is plotted in color code.}
\label{fig:blochoscillation} 
\end{figure}

The quasienergy band structure of effective Hamiltonian $H_{eff}$ in different cases are shown in Fig.\ref{fig:blochoscillation}(a-c).  The central bands are flat for all three cases that: (i) $\kappa_1 = \kappa_2$ with $A_0=0$, the Floquet system is a semi-metal with a spin-1 Dirac cone in Fig.\ref{fig:blochoscillation}(a);  (ii) $\kappa_1 \ne \kappa_2$ with $A_0 = 0$, a trivial insulator in Fig.\ref{fig:blochoscillation}(b); and (iii) $A_0 \ne 0$ with $\kappa_1 = \kappa_2$, the system becomes a Floquet topological insulator\cite{weeks2010topological,bandres2014lieb}. The topology of each band is indicated by the integer Chern number $C_i=\frac{i}{2\pi}\oint \langle u_{i, \bm{k}} |\nabla_{\bm{k}}|u_{i,\bm{k}}\rangle d\bm{k}$, $(C_i\in 0,\pm 1)$, where $u_{i,\bm{k}}$ is the eigenstate of $i$-th band. From the calculated band structure, for these cases of Fig.~\ref{fig:blochoscillation} (a-c), the band structures appear to be invariant under either the particle-hole symmetry operation $\beta \to -\beta $ or time reversal system $k \to -k$, even with a non-zero effective gauge field $\bm{A}(z)$  (More details in Supplement). To apply these ideas to a realistic photonic lattice, such as femtosecond laser-written waveguides in fused silica~\cite{guzman2014experimental,vicencio2015observation,mukherjee2015observation,diebel2016conical}, the refractive index is $n_0 = 1.45$ at wavelength $\lambda = 633 nm$, with modulation $\Delta n = 7.0 \times 10^{-4}$ for beam propagation equation Eq.~(\ref{eq:schrodingerequation}). According to real experiments, the waveguides elliptical cross sections have axis diameters 11$\mu m$ and 4 $\mu m$, as shown in Fig.~\ref{fig:floquetlieblattice}(b). They form a Floquet Lieb lattice with the lattice constant $a$, helix radius $R$, and pitch $Z_0$, where the effective periodic-driven gauge field can be adjusted by setting appropriate $a$, $R$, and $Z_0$.

\vskip0.25cm\noindent
{\bf Topological Landau-Zener Bloch Oscillations in Floquet Lieb lattices.}
We now consider the case that an effective fore field $\bm{F} = (F_x,F_y)^T$ is applied to provide a source for Bloch oscillations. This effective force field in continuum photonic medium can be attained from gradient index, which can be induced optically and adjusted by the fused power\cite{trompeter2006bloch}  (More details in Supplement). The $\Gamma$ point flat band state is chosen as the initial state, excited by a Gaussian wave packet at the center of Lieb lattice with on-site weight along $M$ direction and periodic along its perpendicular direction, shown in Fig.~\ref{fig:blochoscillation}(g). For simplicity, we consider the case that the effective force field is along $M$ direction,  $F_x=F_y=F$ and the following discussion will focus on the tight-binding model for computational convenience. 

Fig.~\ref{fig:blochoscillation}(d-e) display the numerical results of Bloch oscillations. In these cases that $A_0 =0$ with a flat band in Fig.~\ref{fig:blochoscillation}(d,e), no oscillations occur. For the case that $\kappa_1 = \kappa_2$ and $A_0 \ne 0$ in Fig.~\ref{fig:blochoscillation}(f), the initial state starts to evolve in a symmetric way, reflecting the band structure symmetry in Fig.~\ref{fig:blochoscillation}(a).  For these cases with $A_0 =0$, it should be mentioned that there are no Landau-Zener transitions observed. The flat band state in these settings is still localized state. It behaves like the ``dark'' state which can not couple with other eigen-states even during non-adiabatic process. But the situation for flat band will be different in topological insulator. For the topological insulator with the band structure in the Fig.~\ref{fig:blochoscillation}(c), bands can be associated with nonzero topological invariants, i.e, Chern number, and there is no isolated band, even for flat band that is seemingly separated by the topological gap. This is more transparent if we look into the band structures of a semi-infinite Floquet Lieb lattice strip with edges. Topological edge state emerges, across the topological nontrivial gap, connecting the different bulk bands. The unidirectional edge transport is robust to the disorder and perturbation (More details in Supplement).

The topological transition is one kind of global properties in momentum space for topological insulators. Thus the Landau-Zener tunneling from flat band to other bands by crossing the topological nontrivial gap can be directly observed during the Bloch oscillation with non-adiabatic process, see Fig.~\ref{fig:blochoscillation}(f).
In particular, for the Bloch oscillation shown in Fig.~\ref{fig:blochoscillation}(f), we observed the constructive and destructive interference of wave packets, whose behaviors can be described by the Heisenberg-like equation of motion (More details in Supplement):
\begin{equation}
i\frac{\partial}{\partial z}\vec A = \hat{\beta}_{\bm{k}} \vec A + F x_0 \vec A + F \hat \Theta \vec{1}
\end{equation}
where $A = (A_1(z),A_2(z),A_3(z))^T $ represents the components of three bands during Bloch oscillations and $\hat{\beta}_{\bm{k}} = diag(\beta_{1,\bm{k}},\beta_{2,\bm{k}},\beta_{3,\bm{k}}) $, $\beta_{n,\bm{k}}$ is the energy of $n$-th band, $\vec{1}=(1,1,1)^T$. The matrix $\hat \Theta$ reveal the all interactions among three bands and has the elements:
\begin{equation}
\Theta_{ij}=i \langle u_{i,\bm{k}}|\partial_{\bm{k}} u_{j,\bm{k}}\rangle,
\end{equation}
The diagonal terms of $\Theta_{ii}$, reminiscent of the Berry curvature, provide the geometrical phase accumulated during Bloch oscillation. The off-diagonal terms describe the non-adiabatic Landau-Zener tunneling process between different bands.  It is clear that for the Bloch oscillation, the phases accumulated during the process have four parts of contributions\cite{shevchenko2010landau}: dynamic phase $\varphi_{dyn}$, Zeeman phase $\varphi_{Zeeman}$ contributed by external field ,  Stokes phase $\varphi_{S}$ raised from non-adiabatic process\cite{kayanuma1997stokes}, and importantly the geometric phase $\varphi_{g}$ came from the topological phase transition. The sum of these four phases leads to the collective interference behavior for Bloch oscillations (More details in Supplymentary):
\begin{equation}
\varphi_{total} = \varphi_{dyn} + \varphi_{Zeeman} + \varphi_{S}  + \varphi_{g}
\label{eq:phasesum}
\end{equation}
In these phases, $\varphi_{dyn}$ is determined by the band structure, there are no remarkable differences for four cases in Fig.\ref{fig:blochoscillation}. $\varphi_{Zeeman}$ can be cancelled by tuning external effective force field, and $\varphi_{S}$ is only associated with the non-adiabatic Landau-Zener tunneling process, around the point $M$ in the momentum space. But for the geometric phase $\varphi_{g}$, it becomes non-trivial only after topological phase transition. 

\begin{figure}[htbp]
\centering
\includegraphics[width=\linewidth]{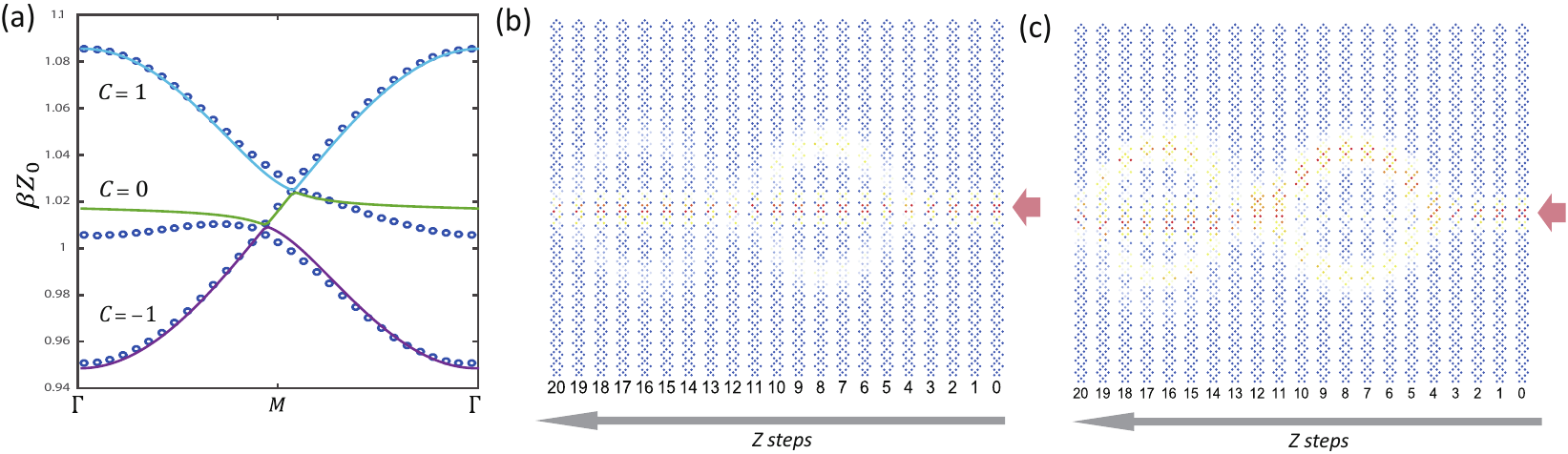}
\caption{
{\bf Dimertized topological Lieb lattice and the effect raised from topological transition to Landau-Zener Bloch oscillation.} (a) Quasi-energy bandstructure of dimertized topological Lieb lattice of helical waveguides($\kappa_1\ne \kappa_2$, $A_0\ne0$). (b)Topological Trivial case before the transition. (c) Topological non-trivial case. After topological transition, the topological phase accumulated in the adiabatic process of Bloch oscillation can interfere with the original phases, such as dynamic phase, Zeeman phase and Stokes phase. The Bloch oscillation of period $5Z_0$ is displayed.}
\label{fig:landauzenerblochoscillation}
\end{figure}

Let us scrutinize the difference of Bloch oscillations before and after the topological transition. For the case in Fig.\ref{fig:blochoscillation}(c), two topological transitions emerge from the triple degeneracy at the same momentum point $M$ in Fig.\ref{fig:blochoscillation}(a). The topological transition positions of two gaps between the central flat band and the upper or lower bands are thus indistinguishable.  To clarify the effect of topological transitions on Bloch oscillation, we use the dimerized Floquet Lieb lattice  to separate and observe individually the transition points.
As shown in Fig.\ref{fig:landauzenerblochoscillation}(a), with dimerized nearest-neighbor coupling $\kappa_1 \ne \kappa_2$, the central flat band becomes perturbed and two gaps symmetrically locate around the $M$ and the central ``zero'' energy. Since dimertized nearest-neighbor coupling leads to lower spectral symmetry of the general case, the spectrum becomes invariant under the combined action of both particle-hole symmetry and time reversal operation, namely $\beta \to -\beta$ and $k \to -k$.  To achieve dimertized nearest-neigbhor coupling $\kappa_1 \ne \kappa_2$, the distant between nearest waveguides along $x$ and $y$ direction will be different, i.e, $d_1 \ne d_2$, shown in Fig.\ref{fig:floquetlieblattice}(b).

\begin{figure}[htbp]
\centering
\includegraphics[width=\linewidth]{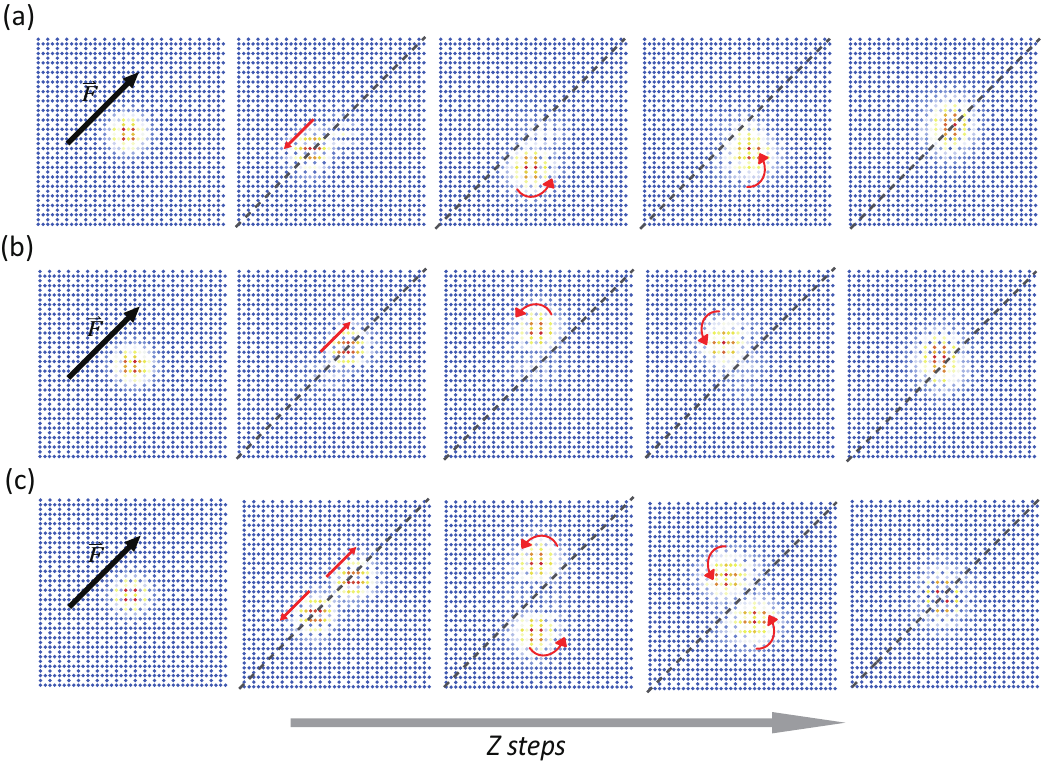}
\caption{{\bf Anomalous velocities in Bloch oscillations of Floquet topological Lieb insulators of helical waveguides.} The anomalous velocities result from the non-trivial Berry curvatures for each Floquet band. Here we adjust the strength of effective gauge field $A_0$ large enough to open wide non-trivial gap for avoiding non-adiabatic process.The initial state is excited by a Gaussian wave pocket with eigenstate of upper band(a), lower band(b), and the linear superposition of two states(c), respectively.}
\label{fig:anomalousvelocities}
\end{figure}

To realize topological transition from trivial insulator into non-trivial one in Fig.\ref{fig:landauzenerblochoscillation}(b,c), we adjust the strength of effective gauge field $A_0$ to achieve this goal in the dimertized Lieb lattice, and then the topological trivial gap will close and reopen a non-trivial gap. In the results from Fig.\ref{fig:landauzenerblochoscillation}, there are two cases: (i) topological trivial gap, (ii) topological non-trivial gap. We observed the main difference between trivial and non-trivial cases: the disappearance of central flat band state. The mechanism behind this phenomenon comes from the destructive interference after involving in the effect of geometrical phase. For classical Bloch oscillation, St\"uckelberg phases are responsible for constructive and destructive interference of wave functions, but for topological insulators, the sum of St\"uckelberg phases and geometrical phases will play the main role. Specially, the distribution of geometrical phases in momentum space is not uniform for two-dimensional photonics topological insulator, which means the Bloch oscillation is angle-dependent. For the system of Fig.\ref{fig:landauzenerblochoscillation}(a) we discuss here, $M$ and $M'$ are two distinct points in momentum space.

Another unique property in topological Lieb lattices is the anomalous perpendicular velocity\cite{xiao2010berry} to the direction of the effective force field $\bm{F}$. This anomalous velocity comes from the non-zero Berry curvature of Floquet bands. For the state $|u_{n, \bm{k}}\rangle$ of the $n$-th band, the anomalous distribution of transverse motion reads $v = - F \Omega_n(\bm{k})$, where $\Omega_n = \Theta_{nn}=i \langle u_{n,\bm{k}}|\partial_{\bm{k}} u_{n,\bm{k}} \rangle$ is the Berry curvature of the $n$-th band. To ignore the effect of non-adiabatic process, we adjust the strength of effective gauge field $A_0$ to a large value and observe that there exist anomalous velocities during Bloch oscillation shown in Fig.\ref{fig:anomalousvelocities}. Two states of upper and lower bands, and their linear superposition are used to excite the Bloch oscillation. From the numerical results, we can find that the evolutions of states are along circular shape motion trajectories, which are the combined effects of external effective force field and effective non-zero gauge field. The anomalous velocities can be used to reveal the distribution of Berry curvature in Brillouin zone and then measure the topological invariants of bands, i.e, Chern number\cite{aidelsburger2015measuring}. 

\vskip0.25cm\noindent
{\bf Conclusion}\\
In summary, we have investigated the mechanism of topological Landau-Zener Bloch oscillation in  photonic Floquet Lieb insulators. The artificial gauge field or synthetic spin-orbital interaction of light has been achieved by the so-called Floquet mechanism, the periodic helical twisting of  photonic waveguides.
The localized states of flat bands in photonic Floquet Lieb insulators can be entangled with other bands through the non-trivial band topology, characterized by nonzero topological invariants, i.e, Chern number. The geometrical phases raised from topological transition can make a complicated interference with the conventional phase of Landau-Zener Bloch oscillation. 
The combination of the physics of Bloch oscillations, Landau-Zener tunneling, flat band, and topological insulators opens promising directions for controlling unconventional quantum state and transport through properly designed lattices and interactions.
The Lieb Lattice has exhibited intriguing properties that are of great interest in both the fundamental physics and practical applications~\cite{Drost2017, slot2017experimental}. The many interesting behaviors of the topological phases in Floquet Lieb lattices in Boson-Einstein condensate\cite{ozawa2017interaction}, chiral bound states\cite{mur2014chiral}, and Zeno Hall effect\cite{gong2017zeno}, are worth exploring in detail for further work.

\bibliography{references}

\begin{thebibliography}{46}%
\makeatletter
\providecommand \@ifxundefined [1]{%
 \@ifx{#1\undefined}
}%
\providecommand \@ifnum [1]{%
 \ifnum #1\expandafter \@firstoftwo
 \else \expandafter \@secondoftwo
 \fi
}%
\providecommand \@ifx [1]{%
 \ifx #1\expandafter \@firstoftwo
 \else \expandafter \@secondoftwo
 \fi
}%
\providecommand \natexlab [1]{#1}%
\providecommand \enquote  [1]{``#1''}%
\providecommand \bibnamefont  [1]{#1}%
\providecommand \bibfnamefont [1]{#1}%
\providecommand \citenamefont [1]{#1}%
\providecommand \href@noop [0]{\@secondoftwo}%
\providecommand \href [0]{\begingroup \@sanitize@url \@href}%
\providecommand \@href[1]{\@@startlink{#1}\@@href}%
\providecommand \@@href[1]{\endgroup#1\@@endlink}%
\providecommand \@sanitize@url [0]{\catcode `\\12\catcode `\$12\catcode
  `\&12\catcode `\#12\catcode `\^12\catcode `\_12\catcode `\%12\relax}%
\providecommand \@@startlink[1]{}%
\providecommand \@@endlink[0]{}%
\providecommand \url  [0]{\begingroup\@sanitize@url \@url }%
\providecommand \@url [1]{\endgroup\@href {#1}{\urlprefix }}%
\providecommand \urlprefix  [0]{URL }%
\providecommand \Eprint [0]{\href }%
\providecommand \doibase [0]{http://dx.doi.org/}%
\providecommand \selectlanguage [0]{\@gobble}%
\providecommand \bibinfo  [0]{\@secondoftwo}%
\providecommand \bibfield  [0]{\@secondoftwo}%
\providecommand \translation [1]{[#1]}%
\providecommand \BibitemOpen [0]{}%
\providecommand \bibitemStop [0]{}%
\providecommand \bibitemNoStop [0]{.\EOS\space}%
\providecommand \EOS [0]{\spacefactor3000\relax}%
\providecommand \BibitemShut  [1]{\csname bibitem#1\endcsname}%
\let\auto@bib@innerbib\@empty
\bibitem [{\citenamefont {Moore}(2010)}]{Moore2010The}%
  \BibitemOpen
  \bibfield  {author} {\bibinfo {author} {\bibfnamefont {J.~E.}\ \bibnamefont
  {Moore}},\ }\href@noop {} {\bibfield  {journal} {\bibinfo  {journal}
  {Nature}\ }\textbf {\bibinfo {volume} {464}},\ \bibinfo {pages} {194}
  (\bibinfo {year} {2010})}\BibitemShut {NoStop}%
\bibitem [{\citenamefont {Hasan}\ and\ \citenamefont
  {Kane}(2010)}]{Hasan2010Colloquium}%
  \BibitemOpen
  \bibfield  {author} {\bibinfo {author} {\bibfnamefont {M.~Z.}\ \bibnamefont
  {Hasan}}\ and\ \bibinfo {author} {\bibfnamefont {C.~L.}\ \bibnamefont
  {Kane}},\ }\href@noop {} {\bibfield  {journal} {\bibinfo  {journal} {Reviews
  of Modern Physics}\ }\textbf {\bibinfo {volume} {82}},\ \bibinfo {pages}
  {3045} (\bibinfo {year} {2010})}\BibitemShut {NoStop}%
\bibitem [{\citenamefont {Qi}\ and\ \citenamefont
  {Zhang}(2011)}]{qi2011topological}%
  \BibitemOpen
  \bibfield  {author} {\bibinfo {author} {\bibfnamefont {X.-L.}\ \bibnamefont
  {Qi}}\ and\ \bibinfo {author} {\bibfnamefont {S.-C.}\ \bibnamefont {Zhang}},\
  }\href@noop {} {\bibfield  {journal} {\bibinfo  {journal} {Reviews of Modern
  Physics}\ }\textbf {\bibinfo {volume} {83}},\ \bibinfo {pages} {1057}
  (\bibinfo {year} {2011})}\BibitemShut {NoStop}%
\bibitem [{\citenamefont {Qi}\ \emph {et~al.}(2009)\citenamefont {Qi},
  \citenamefont {Li}, \citenamefont {Zang},\ and\ \citenamefont
  {Zhang}}]{Qi2009Inducing}%
  \BibitemOpen
  \bibfield  {author} {\bibinfo {author} {\bibfnamefont {X.~L.}\ \bibnamefont
  {Qi}}, \bibinfo {author} {\bibfnamefont {R.}~\bibnamefont {Li}}, \bibinfo
  {author} {\bibfnamefont {J.}~\bibnamefont {Zang}}, \ and\ \bibinfo {author}
  {\bibfnamefont {S.~C.}\ \bibnamefont {Zhang}},\ }\href@noop {} {\bibfield
  {journal} {\bibinfo  {journal} {Science}\ }\textbf {\bibinfo {volume}
  {323}},\ \bibinfo {pages} {1184} (\bibinfo {year} {2009})}\BibitemShut
  {NoStop}%
\bibitem [{\citenamefont {Xu}\ \emph {et~al.}(2015)\citenamefont {Xu},
  \citenamefont {Belopolski}, \citenamefont {Alidoust}, \citenamefont
  {Neupane}, \citenamefont {Bian}, \citenamefont {Zhang}, \citenamefont
  {Sankar}, \citenamefont {Chang}, \citenamefont {Yuan},\ and\ \citenamefont
  {Lee}}]{Xu2015TOPOLOGICAL}%
  \BibitemOpen
  \bibfield  {author} {\bibinfo {author} {\bibfnamefont {S.~Y.}\ \bibnamefont
  {Xu}}, \bibinfo {author} {\bibfnamefont {I.}~\bibnamefont {Belopolski}},
  \bibinfo {author} {\bibfnamefont {N.}~\bibnamefont {Alidoust}}, \bibinfo
  {author} {\bibfnamefont {M.}~\bibnamefont {Neupane}}, \bibinfo {author}
  {\bibfnamefont {G.}~\bibnamefont {Bian}}, \bibinfo {author} {\bibfnamefont
  {C.}~\bibnamefont {Zhang}}, \bibinfo {author} {\bibfnamefont
  {R.}~\bibnamefont {Sankar}}, \bibinfo {author} {\bibfnamefont
  {G.}~\bibnamefont {Chang}}, \bibinfo {author} {\bibfnamefont
  {Z.}~\bibnamefont {Yuan}}, \ and\ \bibinfo {author} {\bibfnamefont {C.~C.}\
  \bibnamefont {Lee}},\ }\href@noop {} {\bibfield  {journal} {\bibinfo
  {journal} {Science}\ }\textbf {\bibinfo {volume} {349}},\ \bibinfo {pages}
  {613} (\bibinfo {year} {2015})}\BibitemShut {NoStop}%
\bibitem [{\citenamefont {Rechtsman}\ \emph
  {et~al.}(2013{\natexlab{a}})\citenamefont {Rechtsman}, \citenamefont
  {Zeuner}, \citenamefont {Plotnik}, \citenamefont {Lumer}, \citenamefont
  {Nolte}, \citenamefont {Segev},\ and\ \citenamefont
  {Szameit}}]{Rechtsman2013Photonic}%
  \BibitemOpen
  \bibfield  {author} {\bibinfo {author} {\bibfnamefont {M.}~\bibnamefont
  {Rechtsman}}, \bibinfo {author} {\bibfnamefont {J.}~\bibnamefont {Zeuner}},
  \bibinfo {author} {\bibfnamefont {Y.}~\bibnamefont {Plotnik}}, \bibinfo
  {author} {\bibfnamefont {Y.}~\bibnamefont {Lumer}}, \bibinfo {author}
  {\bibfnamefont {S.}~\bibnamefont {Nolte}}, \bibinfo {author} {\bibfnamefont
  {M.}~\bibnamefont {Segev}}, \ and\ \bibinfo {author} {\bibfnamefont
  {A.}~\bibnamefont {Szameit}},\ }\href@noop {} {\bibfield  {journal} {\bibinfo
   {journal} {Nature Materials}\ }\textbf {\bibinfo {volume} {12}},\ \bibinfo
  {pages} {233} (\bibinfo {year} {2013}{\natexlab{a}})}\BibitemShut {NoStop}%
\bibitem [{\citenamefont {Yang}\ \emph {et~al.}(2015)\citenamefont {Yang},
  \citenamefont {Gao}, \citenamefont {Shi}, \citenamefont {Lin}, \citenamefont
  {Gao}, \citenamefont {Chong},\ and\ \citenamefont
  {Zhang}}]{yang2015topological}%
  \BibitemOpen
  \bibfield  {author} {\bibinfo {author} {\bibfnamefont {Z.}~\bibnamefont
  {Yang}}, \bibinfo {author} {\bibfnamefont {F.}~\bibnamefont {Gao}}, \bibinfo
  {author} {\bibfnamefont {X.}~\bibnamefont {Shi}}, \bibinfo {author}
  {\bibfnamefont {X.}~\bibnamefont {Lin}}, \bibinfo {author} {\bibfnamefont
  {Z.}~\bibnamefont {Gao}}, \bibinfo {author} {\bibfnamefont {Y.}~\bibnamefont
  {Chong}}, \ and\ \bibinfo {author} {\bibfnamefont {B.}~\bibnamefont
  {Zhang}},\ }\href@noop {} {\bibfield  {journal} {\bibinfo  {journal}
  {Physical review letters}\ }\textbf {\bibinfo {volume} {114}},\ \bibinfo
  {pages} {114301} (\bibinfo {year} {2015})}\BibitemShut {NoStop}%
\bibitem [{\citenamefont {Lu}\ \emph {et~al.}(2014)\citenamefont {Lu},
  \citenamefont {Joannopoulos},\ and\ \citenamefont {Solja{\v
  c}i{\'c}}}]{Lu2014Topological}%
  \BibitemOpen
  \bibfield  {author} {\bibinfo {author} {\bibfnamefont {L.}~\bibnamefont
  {Lu}}, \bibinfo {author} {\bibfnamefont {J.~D.}\ \bibnamefont
  {Joannopoulos}}, \ and\ \bibinfo {author} {\bibfnamefont {M.}~\bibnamefont
  {Solja{\v c}i{\'c}}},\ }\href@noop {} {\bibfield  {journal} {\bibinfo
  {journal} {Nature Photonics}\ }\textbf {\bibinfo {volume} {8}},\ \bibinfo
  {pages} {821} (\bibinfo {year} {2014})}\BibitemShut {NoStop}%
\bibitem [{\citenamefont {Liang}\ and\ \citenamefont
  {Chong}(2012)}]{Liang2012Optical}%
  \BibitemOpen
  \bibfield  {author} {\bibinfo {author} {\bibfnamefont {G.~Q.}\ \bibnamefont
  {Liang}}\ and\ \bibinfo {author} {\bibfnamefont {Y.~D.}\ \bibnamefont
  {Chong}},\ }\href@noop {} {\bibfield  {journal} {\bibinfo  {journal}
  {Physical Review Letters}\ }\textbf {\bibinfo {volume} {110}},\ \bibinfo
  {pages} {306} (\bibinfo {year} {2012})}\BibitemShut {NoStop}%
\bibitem [{\citenamefont {Ningyuan}\ \emph {et~al.}(2015)\citenamefont
  {Ningyuan}, \citenamefont {Owens}, \citenamefont {Sommer}, \citenamefont
  {Schuster},\ and\ \citenamefont {Simon}}]{ningyuan2015time}%
  \BibitemOpen
  \bibfield  {author} {\bibinfo {author} {\bibfnamefont {J.}~\bibnamefont
  {Ningyuan}}, \bibinfo {author} {\bibfnamefont {C.}~\bibnamefont {Owens}},
  \bibinfo {author} {\bibfnamefont {A.}~\bibnamefont {Sommer}}, \bibinfo
  {author} {\bibfnamefont {D.}~\bibnamefont {Schuster}}, \ and\ \bibinfo
  {author} {\bibfnamefont {J.}~\bibnamefont {Simon}},\ }\href@noop {}
  {\bibfield  {journal} {\bibinfo  {journal} {Physical Review X}\ }\textbf
  {\bibinfo {volume} {5}},\ \bibinfo {pages} {021031} (\bibinfo {year}
  {2015})}\BibitemShut {NoStop}%
\bibitem [{\citenamefont {Wu}\ and\ \citenamefont {Hu}(2015)}]{wu2015scheme}%
  \BibitemOpen
  \bibfield  {author} {\bibinfo {author} {\bibfnamefont {L.-H.}\ \bibnamefont
  {Wu}}\ and\ \bibinfo {author} {\bibfnamefont {X.}~\bibnamefont {Hu}},\
  }\href@noop {} {\bibfield  {journal} {\bibinfo  {journal} {Physical review
  letters}\ }\textbf {\bibinfo {volume} {114}},\ \bibinfo {pages} {223901}
  (\bibinfo {year} {2015})}\BibitemShut {NoStop}%
\bibitem [{\citenamefont {Gao}\ \emph {et~al.}(2015)\citenamefont {Gao},
  \citenamefont {Lawrence}, \citenamefont {Yang}, \citenamefont {Liu},
  \citenamefont {Fang}, \citenamefont {B{\'e}ri}, \citenamefont {Li},\ and\
  \citenamefont {Zhang}}]{gao2015topological}%
  \BibitemOpen
  \bibfield  {author} {\bibinfo {author} {\bibfnamefont {W.}~\bibnamefont
  {Gao}}, \bibinfo {author} {\bibfnamefont {M.}~\bibnamefont {Lawrence}},
  \bibinfo {author} {\bibfnamefont {B.}~\bibnamefont {Yang}}, \bibinfo {author}
  {\bibfnamefont {F.}~\bibnamefont {Liu}}, \bibinfo {author} {\bibfnamefont
  {F.}~\bibnamefont {Fang}}, \bibinfo {author} {\bibfnamefont {B.}~\bibnamefont
  {B{\'e}ri}}, \bibinfo {author} {\bibfnamefont {J.}~\bibnamefont {Li}}, \ and\
  \bibinfo {author} {\bibfnamefont {S.}~\bibnamefont {Zhang}},\ }\href@noop {}
  {\bibfield  {journal} {\bibinfo  {journal} {Physical review letters}\
  }\textbf {\bibinfo {volume} {114}},\ \bibinfo {pages} {037402} (\bibinfo
  {year} {2015})}\BibitemShut {NoStop}%
\bibitem [{\citenamefont {Rechtsman}\ \emph
  {et~al.}(2013{\natexlab{b}})\citenamefont {Rechtsman}, \citenamefont
  {Zeuner}, \citenamefont {Plotnik}, \citenamefont {Lumer}, \citenamefont
  {Podolsky}, \citenamefont {Dreisow}, \citenamefont {Nolte}, \citenamefont
  {Segev},\ and\ \citenamefont {Szameit}}]{rechtsman2013floquet}%
  \BibitemOpen
  \bibfield  {author} {\bibinfo {author} {\bibfnamefont {M.~C.}\ \bibnamefont
  {Rechtsman}}, \bibinfo {author} {\bibfnamefont {J.~M.}\ \bibnamefont
  {Zeuner}}, \bibinfo {author} {\bibfnamefont {Y.}~\bibnamefont {Plotnik}},
  \bibinfo {author} {\bibfnamefont {Y.}~\bibnamefont {Lumer}}, \bibinfo
  {author} {\bibfnamefont {D.}~\bibnamefont {Podolsky}}, \bibinfo {author}
  {\bibfnamefont {F.}~\bibnamefont {Dreisow}}, \bibinfo {author} {\bibfnamefont
  {S.}~\bibnamefont {Nolte}}, \bibinfo {author} {\bibfnamefont
  {M.}~\bibnamefont {Segev}}, \ and\ \bibinfo {author} {\bibfnamefont
  {A.}~\bibnamefont {Szameit}},\ }\href@noop {} {\bibfield  {journal} {\bibinfo
   {journal} {Nature}\ }\textbf {\bibinfo {volume} {496}},\ \bibinfo {pages}
  {196} (\bibinfo {year} {2013}{\natexlab{b}})}\BibitemShut {NoStop}%
\bibitem [{\citenamefont {Kitagawa}\ \emph {et~al.}(2010)\citenamefont
  {Kitagawa}, \citenamefont {Berg}, \citenamefont {Rudner},\ and\ \citenamefont
  {Demler}}]{kitagawa2010topological}%
  \BibitemOpen
  \bibfield  {author} {\bibinfo {author} {\bibfnamefont {T.}~\bibnamefont
  {Kitagawa}}, \bibinfo {author} {\bibfnamefont {E.}~\bibnamefont {Berg}},
  \bibinfo {author} {\bibfnamefont {M.}~\bibnamefont {Rudner}}, \ and\ \bibinfo
  {author} {\bibfnamefont {E.}~\bibnamefont {Demler}},\ }\href@noop {}
  {\bibfield  {journal} {\bibinfo  {journal} {Physical Review B}\ }\textbf
  {\bibinfo {volume} {82}},\ \bibinfo {pages} {235114} (\bibinfo {year}
  {2010})}\BibitemShut {NoStop}%
\bibitem [{\citenamefont {Struck}\ \emph {et~al.}(2012)\citenamefont {Struck},
  \citenamefont {{\"O}lschl{\"a}ger}, \citenamefont {Weinberg}, \citenamefont
  {Hauke}, \citenamefont {Simonet}, \citenamefont {Eckardt}, \citenamefont
  {Lewenstein}, \citenamefont {Sengstock},\ and\ \citenamefont
  {Windpassinger}}]{struck2012tunable}%
  \BibitemOpen
  \bibfield  {author} {\bibinfo {author} {\bibfnamefont {J.}~\bibnamefont
  {Struck}}, \bibinfo {author} {\bibfnamefont {C.}~\bibnamefont
  {{\"O}lschl{\"a}ger}}, \bibinfo {author} {\bibfnamefont {M.}~\bibnamefont
  {Weinberg}}, \bibinfo {author} {\bibfnamefont {P.}~\bibnamefont {Hauke}},
  \bibinfo {author} {\bibfnamefont {J.}~\bibnamefont {Simonet}}, \bibinfo
  {author} {\bibfnamefont {A.}~\bibnamefont {Eckardt}}, \bibinfo {author}
  {\bibfnamefont {M.}~\bibnamefont {Lewenstein}}, \bibinfo {author}
  {\bibfnamefont {K.}~\bibnamefont {Sengstock}}, \ and\ \bibinfo {author}
  {\bibfnamefont {P.}~\bibnamefont {Windpassinger}},\ }\href@noop {} {\bibfield
   {journal} {\bibinfo  {journal} {Physical review letters}\ }\textbf {\bibinfo
  {volume} {108}},\ \bibinfo {pages} {225304} (\bibinfo {year}
  {2012})}\BibitemShut {NoStop}%
\bibitem [{\citenamefont {Leykam}\ \emph {et~al.}(2016)\citenamefont {Leykam},
  \citenamefont {Rechtsman},\ and\ \citenamefont
  {Chong}}]{leykam2016anomalous}%
  \BibitemOpen
  \bibfield  {author} {\bibinfo {author} {\bibfnamefont {D.}~\bibnamefont
  {Leykam}}, \bibinfo {author} {\bibfnamefont {M.}~\bibnamefont {Rechtsman}}, \
  and\ \bibinfo {author} {\bibfnamefont {Y.}~\bibnamefont {Chong}},\
  }\href@noop {} {\bibfield  {journal} {\bibinfo  {journal} {Physical Review
  Letters}\ }\textbf {\bibinfo {volume} {117}},\ \bibinfo {pages} {013902}
  (\bibinfo {year} {2016})}\BibitemShut {NoStop}%
\bibitem [{\citenamefont {Guzm{\'a}n-Silva}\ \emph {et~al.}(2014)\citenamefont
  {Guzm{\'a}n-Silva}, \citenamefont {Mej{\'\i}a-Cort{\'e}s}, \citenamefont
  {Bandres}, \citenamefont {Rechtsman}, \citenamefont {Weimann}, \citenamefont
  {Nolte}, \citenamefont {Segev}, \citenamefont {Szameit},\ and\ \citenamefont
  {Vicencio}}]{guzman2014experimental}%
  \BibitemOpen
  \bibfield  {author} {\bibinfo {author} {\bibfnamefont {D.}~\bibnamefont
  {Guzm{\'a}n-Silva}}, \bibinfo {author} {\bibfnamefont {C.}~\bibnamefont
  {Mej{\'\i}a-Cort{\'e}s}}, \bibinfo {author} {\bibfnamefont {M.}~\bibnamefont
  {Bandres}}, \bibinfo {author} {\bibfnamefont {M.}~\bibnamefont {Rechtsman}},
  \bibinfo {author} {\bibfnamefont {S.}~\bibnamefont {Weimann}}, \bibinfo
  {author} {\bibfnamefont {S.}~\bibnamefont {Nolte}}, \bibinfo {author}
  {\bibfnamefont {M.}~\bibnamefont {Segev}}, \bibinfo {author} {\bibfnamefont
  {A.}~\bibnamefont {Szameit}}, \ and\ \bibinfo {author} {\bibfnamefont
  {R.}~\bibnamefont {Vicencio}},\ }\href@noop {} {\bibfield  {journal}
  {\bibinfo  {journal} {New Journal of Physics}\ }\textbf {\bibinfo {volume}
  {16}},\ \bibinfo {pages} {063061} (\bibinfo {year} {2014})}\BibitemShut
  {NoStop}%
\bibitem [{\citenamefont {Vicencio}\ \emph {et~al.}(2015)\citenamefont
  {Vicencio}, \citenamefont {Cantillano}, \citenamefont {Morales-Inostroza},
  \citenamefont {Real}, \citenamefont {Mej{\'\i}a-Cort{\'e}s}, \citenamefont
  {Weimann}, \citenamefont {Szameit},\ and\ \citenamefont
  {Molina}}]{vicencio2015observation}%
  \BibitemOpen
  \bibfield  {author} {\bibinfo {author} {\bibfnamefont {R.~A.}\ \bibnamefont
  {Vicencio}}, \bibinfo {author} {\bibfnamefont {C.}~\bibnamefont
  {Cantillano}}, \bibinfo {author} {\bibfnamefont {L.}~\bibnamefont
  {Morales-Inostroza}}, \bibinfo {author} {\bibfnamefont {B.}~\bibnamefont
  {Real}}, \bibinfo {author} {\bibfnamefont {C.}~\bibnamefont
  {Mej{\'\i}a-Cort{\'e}s}}, \bibinfo {author} {\bibfnamefont {S.}~\bibnamefont
  {Weimann}}, \bibinfo {author} {\bibfnamefont {A.}~\bibnamefont {Szameit}}, \
  and\ \bibinfo {author} {\bibfnamefont {M.~I.}\ \bibnamefont {Molina}},\
  }\href@noop {} {\bibfield  {journal} {\bibinfo  {journal} {Physical review
  letters}\ }\textbf {\bibinfo {volume} {114}},\ \bibinfo {pages} {245503}
  (\bibinfo {year} {2015})}\BibitemShut {NoStop}%
\bibitem [{\citenamefont {Mukherjee}\ \emph {et~al.}(2015)\citenamefont
  {Mukherjee}, \citenamefont {Spracklen}, \citenamefont {Choudhury},
  \citenamefont {Goldman}, \citenamefont {{\"O}hberg}, \citenamefont
  {Andersson},\ and\ \citenamefont {Thomson}}]{mukherjee2015observation}%
  \BibitemOpen
  \bibfield  {author} {\bibinfo {author} {\bibfnamefont {S.}~\bibnamefont
  {Mukherjee}}, \bibinfo {author} {\bibfnamefont {A.}~\bibnamefont
  {Spracklen}}, \bibinfo {author} {\bibfnamefont {D.}~\bibnamefont
  {Choudhury}}, \bibinfo {author} {\bibfnamefont {N.}~\bibnamefont {Goldman}},
  \bibinfo {author} {\bibfnamefont {P.}~\bibnamefont {{\"O}hberg}}, \bibinfo
  {author} {\bibfnamefont {E.}~\bibnamefont {Andersson}}, \ and\ \bibinfo
  {author} {\bibfnamefont {R.~R.}\ \bibnamefont {Thomson}},\ }\href@noop {}
  {\bibfield  {journal} {\bibinfo  {journal} {Physical review letters}\
  }\textbf {\bibinfo {volume} {114}},\ \bibinfo {pages} {245504} (\bibinfo
  {year} {2015})}\BibitemShut {NoStop}%
\bibitem [{\citenamefont {Diebel}\ \emph {et~al.}(2016)\citenamefont {Diebel},
  \citenamefont {Leykam}, \citenamefont {Kroesen}, \citenamefont {Denz},\ and\
  \citenamefont {Desyatnikov}}]{diebel2016conical}%
  \BibitemOpen
  \bibfield  {author} {\bibinfo {author} {\bibfnamefont {F.}~\bibnamefont
  {Diebel}}, \bibinfo {author} {\bibfnamefont {D.}~\bibnamefont {Leykam}},
  \bibinfo {author} {\bibfnamefont {S.}~\bibnamefont {Kroesen}}, \bibinfo
  {author} {\bibfnamefont {C.}~\bibnamefont {Denz}}, \ and\ \bibinfo {author}
  {\bibfnamefont {A.~S.}\ \bibnamefont {Desyatnikov}},\ }\href@noop {}
  {\bibfield  {journal} {\bibinfo  {journal} {Physical review letters}\
  }\textbf {\bibinfo {volume} {116}},\ \bibinfo {pages} {183902} (\bibinfo
  {year} {2016})}\BibitemShut {NoStop}%
\bibitem [{\citenamefont {Mielke}(1999)}]{mielke1999ferromagnetism}%
  \BibitemOpen
  \bibfield  {author} {\bibinfo {author} {\bibfnamefont {A.}~\bibnamefont
  {Mielke}},\ }\href@noop {} {\bibfield  {journal} {\bibinfo  {journal}
  {Physical review letters}\ }\textbf {\bibinfo {volume} {82}},\ \bibinfo
  {pages} {4312} (\bibinfo {year} {1999})}\BibitemShut {NoStop}%
\bibitem [{\citenamefont {Neupert}\ \emph {et~al.}(2011)\citenamefont
  {Neupert}, \citenamefont {Santos}, \citenamefont {Chamon},\ and\
  \citenamefont {Mudry}}]{neupert2011fractional}%
  \BibitemOpen
  \bibfield  {author} {\bibinfo {author} {\bibfnamefont {T.}~\bibnamefont
  {Neupert}}, \bibinfo {author} {\bibfnamefont {L.}~\bibnamefont {Santos}},
  \bibinfo {author} {\bibfnamefont {C.}~\bibnamefont {Chamon}}, \ and\ \bibinfo
  {author} {\bibfnamefont {C.}~\bibnamefont {Mudry}},\ }\href@noop {}
  {\bibfield  {journal} {\bibinfo  {journal} {Physical review letters}\
  }\textbf {\bibinfo {volume} {106}},\ \bibinfo {pages} {236804} (\bibinfo
  {year} {2011})}\BibitemShut {NoStop}%
\bibitem [{\citenamefont {Liu}\ \emph {et~al.}(2013)\citenamefont {Liu},
  \citenamefont {Wang}, \citenamefont {Mei}, \citenamefont {Wu},\ and\
  \citenamefont {Liu}}]{liu2013flat}%
  \BibitemOpen
  \bibfield  {author} {\bibinfo {author} {\bibfnamefont {Z.}~\bibnamefont
  {Liu}}, \bibinfo {author} {\bibfnamefont {Z.-F.}\ \bibnamefont {Wang}},
  \bibinfo {author} {\bibfnamefont {J.-W.}\ \bibnamefont {Mei}}, \bibinfo
  {author} {\bibfnamefont {Y.-S.}\ \bibnamefont {Wu}}, \ and\ \bibinfo {author}
  {\bibfnamefont {F.}~\bibnamefont {Liu}},\ }\href@noop {} {\bibfield
  {journal} {\bibinfo  {journal} {Physical review letters}\ }\textbf {\bibinfo
  {volume} {110}},\ \bibinfo {pages} {106804} (\bibinfo {year}
  {2013})}\BibitemShut {NoStop}%
\bibitem [{\citenamefont {Flach}\ \emph {et~al.}(2014)\citenamefont {Flach},
  \citenamefont {Leykam}, \citenamefont {Bodyfelt}, \citenamefont {Matthies},\
  and\ \citenamefont {Desyatnikov}}]{flach2014detangling}%
  \BibitemOpen
  \bibfield  {author} {\bibinfo {author} {\bibfnamefont {S.}~\bibnamefont
  {Flach}}, \bibinfo {author} {\bibfnamefont {D.}~\bibnamefont {Leykam}},
  \bibinfo {author} {\bibfnamefont {J.~D.}\ \bibnamefont {Bodyfelt}}, \bibinfo
  {author} {\bibfnamefont {P.}~\bibnamefont {Matthies}}, \ and\ \bibinfo
  {author} {\bibfnamefont {A.~S.}\ \bibnamefont {Desyatnikov}},\ }\href@noop {}
  {\bibfield  {journal} {\bibinfo  {journal} {EPL (Europhysics Letters)}\
  }\textbf {\bibinfo {volume} {105}},\ \bibinfo {pages} {30001} (\bibinfo
  {year} {2014})}\BibitemShut {NoStop}%
\bibitem [{\citenamefont {Jacqmin}\ \emph {et~al.}(2014)\citenamefont
  {Jacqmin}, \citenamefont {Carusotto}, \citenamefont {Sagnes}, \citenamefont
  {Abbarchi}, \citenamefont {Solnyshkov}, \citenamefont {Malpuech},
  \citenamefont {Galopin}, \citenamefont {Lema{\^\i}tre}, \citenamefont
  {Bloch},\ and\ \citenamefont {Amo}}]{jacqmin2014direct}%
  \BibitemOpen
  \bibfield  {author} {\bibinfo {author} {\bibfnamefont {T.}~\bibnamefont
  {Jacqmin}}, \bibinfo {author} {\bibfnamefont {I.}~\bibnamefont {Carusotto}},
  \bibinfo {author} {\bibfnamefont {I.}~\bibnamefont {Sagnes}}, \bibinfo
  {author} {\bibfnamefont {M.}~\bibnamefont {Abbarchi}}, \bibinfo {author}
  {\bibfnamefont {D.}~\bibnamefont {Solnyshkov}}, \bibinfo {author}
  {\bibfnamefont {G.}~\bibnamefont {Malpuech}}, \bibinfo {author}
  {\bibfnamefont {E.}~\bibnamefont {Galopin}}, \bibinfo {author} {\bibfnamefont
  {A.}~\bibnamefont {Lema{\^\i}tre}}, \bibinfo {author} {\bibfnamefont
  {J.}~\bibnamefont {Bloch}}, \ and\ \bibinfo {author} {\bibfnamefont
  {A.}~\bibnamefont {Amo}},\ }\href@noop {} {\bibfield  {journal} {\bibinfo
  {journal} {Physical review letters}\ }\textbf {\bibinfo {volume} {112}},\
  \bibinfo {pages} {116402} (\bibinfo {year} {2014})}\BibitemShut {NoStop}%
\bibitem [{\citenamefont {Zener}(1932)}]{zener1932non}%
  \BibitemOpen
  \bibfield  {author} {\bibinfo {author} {\bibfnamefont {C.}~\bibnamefont
  {Zener}},\ }in\ \href@noop {} {\emph {\bibinfo {booktitle} {Proceedings of
  the Royal Society of London A: Mathematical, Physical and Engineering
  Sciences}}},\ Vol.\ \bibinfo {volume} {137}\ (\bibinfo {organization} {The
  Royal Society},\ \bibinfo {year} {1932})\ pp.\ \bibinfo {pages}
  {696--702}\BibitemShut {NoStop}%
\bibitem [{\citenamefont {Khomeriki}\ and\ \citenamefont
  {Flach}(2016)}]{khomeriki2016landau}%
  \BibitemOpen
  \bibfield  {author} {\bibinfo {author} {\bibfnamefont {R.}~\bibnamefont
  {Khomeriki}}\ and\ \bibinfo {author} {\bibfnamefont {S.}~\bibnamefont
  {Flach}},\ }\href@noop {} {\bibfield  {journal} {\bibinfo  {journal}
  {Physical Review Letters}\ }\textbf {\bibinfo {volume} {116}},\ \bibinfo
  {pages} {245301} (\bibinfo {year} {2016})}\BibitemShut {NoStop}%
\bibitem [{\citenamefont {Kling}\ \emph {et~al.}(2010)\citenamefont {Kling},
  \citenamefont {Salger}, \citenamefont {Grossert},\ and\ \citenamefont
  {Weitz}}]{kling2010atomic}%
  \BibitemOpen
  \bibfield  {author} {\bibinfo {author} {\bibfnamefont {S.}~\bibnamefont
  {Kling}}, \bibinfo {author} {\bibfnamefont {T.}~\bibnamefont {Salger}},
  \bibinfo {author} {\bibfnamefont {C.}~\bibnamefont {Grossert}}, \ and\
  \bibinfo {author} {\bibfnamefont {M.}~\bibnamefont {Weitz}},\ }\href@noop {}
  {\bibfield  {journal} {\bibinfo  {journal} {Physical review letters}\
  }\textbf {\bibinfo {volume} {105}},\ \bibinfo {pages} {215301} (\bibinfo
  {year} {2010})}\BibitemShut {NoStop}%
\bibitem [{\citenamefont {Morandotti}\ \emph {et~al.}(1999)\citenamefont
  {Morandotti}, \citenamefont {Peschel}, \citenamefont {Aitchison},
  \citenamefont {Eisenberg},\ and\ \citenamefont
  {Silberberg}}]{morandotti1999experimental}%
  \BibitemOpen
  \bibfield  {author} {\bibinfo {author} {\bibfnamefont {R.}~\bibnamefont
  {Morandotti}}, \bibinfo {author} {\bibfnamefont {U.}~\bibnamefont {Peschel}},
  \bibinfo {author} {\bibfnamefont {J.}~\bibnamefont {Aitchison}}, \bibinfo
  {author} {\bibfnamefont {H.}~\bibnamefont {Eisenberg}}, \ and\ \bibinfo
  {author} {\bibfnamefont {Y.}~\bibnamefont {Silberberg}},\ }\href@noop {}
  {\bibfield  {journal} {\bibinfo  {journal} {Physical Review Letters}\
  }\textbf {\bibinfo {volume} {83}},\ \bibinfo {pages} {4756} (\bibinfo {year}
  {1999})}\BibitemShut {NoStop}%
\bibitem [{\citenamefont {Peschel}\ \emph {et~al.}(1998)\citenamefont
  {Peschel}, \citenamefont {Pertsch},\ and\ \citenamefont
  {Lederer}}]{peschel1998optical}%
  \BibitemOpen
  \bibfield  {author} {\bibinfo {author} {\bibfnamefont {U.}~\bibnamefont
  {Peschel}}, \bibinfo {author} {\bibfnamefont {T.}~\bibnamefont {Pertsch}}, \
  and\ \bibinfo {author} {\bibfnamefont {F.}~\bibnamefont {Lederer}},\
  }\href@noop {} {\bibfield  {journal} {\bibinfo  {journal} {Optics letters}\
  }\textbf {\bibinfo {volume} {23}},\ \bibinfo {pages} {1701} (\bibinfo {year}
  {1998})}\BibitemShut {NoStop}%
\bibitem [{\citenamefont {Cayssol}\ \emph {et~al.}(2013)\citenamefont
  {Cayssol}, \citenamefont {D{\'o}ra}, \citenamefont {Simon},\ and\
  \citenamefont {Moessner}}]{cayssol2013floquet}%
  \BibitemOpen
  \bibfield  {author} {\bibinfo {author} {\bibfnamefont {J.}~\bibnamefont
  {Cayssol}}, \bibinfo {author} {\bibfnamefont {B.}~\bibnamefont {D{\'o}ra}},
  \bibinfo {author} {\bibfnamefont {F.}~\bibnamefont {Simon}}, \ and\ \bibinfo
  {author} {\bibfnamefont {R.}~\bibnamefont {Moessner}},\ }\href@noop {}
  {\bibfield  {journal} {\bibinfo  {journal} {physica status solidi (RRL)-Rapid
  Research Letters}\ }\textbf {\bibinfo {volume} {7}},\ \bibinfo {pages} {101}
  (\bibinfo {year} {2013})}\BibitemShut {NoStop}%
\bibitem [{\citenamefont {Weeks}\ and\ \citenamefont
  {Franz}(2010)}]{weeks2010topological}%
  \BibitemOpen
  \bibfield  {author} {\bibinfo {author} {\bibfnamefont {C.}~\bibnamefont
  {Weeks}}\ and\ \bibinfo {author} {\bibfnamefont {M.}~\bibnamefont {Franz}},\
  }\href@noop {} {\bibfield  {journal} {\bibinfo  {journal} {Physical Review
  B}\ }\textbf {\bibinfo {volume} {82}},\ \bibinfo {pages} {085310} (\bibinfo
  {year} {2010})}\BibitemShut {NoStop}%
\bibitem [{\citenamefont {Bandres}\ \emph {et~al.}(2014)\citenamefont
  {Bandres}, \citenamefont {Rechtsman}, \citenamefont {Szameit},\ and\
  \citenamefont {Segev}}]{bandres2014lieb}%
  \BibitemOpen
  \bibfield  {author} {\bibinfo {author} {\bibfnamefont {M.~A.}\ \bibnamefont
  {Bandres}}, \bibinfo {author} {\bibfnamefont {M.~C.}\ \bibnamefont
  {Rechtsman}}, \bibinfo {author} {\bibfnamefont {A.}~\bibnamefont {Szameit}},
  \ and\ \bibinfo {author} {\bibfnamefont {M.}~\bibnamefont {Segev}},\ }in\
  \href@noop {} {\emph {\bibinfo {booktitle} {Lasers and Electro-Optics (CLEO),
  2014 Conference on}}}\ (\bibinfo {organization} {IEEE},\ \bibinfo {year}
  {2014})\ pp.\ \bibinfo {pages} {1--2}\BibitemShut {NoStop}%
\bibitem [{\citenamefont {Trompeter}\ \emph {et~al.}(2006)\citenamefont
  {Trompeter}, \citenamefont {Krolikowski}, \citenamefont {Neshev},
  \citenamefont {Desyatnikov}, \citenamefont {Sukhorukov}, \citenamefont
  {Kivshar}, \citenamefont {Pertsch}, \citenamefont {Peschel},\ and\
  \citenamefont {Lederer}}]{trompeter2006bloch}%
  \BibitemOpen
  \bibfield  {author} {\bibinfo {author} {\bibfnamefont {H.}~\bibnamefont
  {Trompeter}}, \bibinfo {author} {\bibfnamefont {W.}~\bibnamefont
  {Krolikowski}}, \bibinfo {author} {\bibfnamefont {D.~N.}\ \bibnamefont
  {Neshev}}, \bibinfo {author} {\bibfnamefont {A.~S.}\ \bibnamefont
  {Desyatnikov}}, \bibinfo {author} {\bibfnamefont {A.~A.}\ \bibnamefont
  {Sukhorukov}}, \bibinfo {author} {\bibfnamefont {Y.~S.}\ \bibnamefont
  {Kivshar}}, \bibinfo {author} {\bibfnamefont {T.}~\bibnamefont {Pertsch}},
  \bibinfo {author} {\bibfnamefont {U.}~\bibnamefont {Peschel}}, \ and\
  \bibinfo {author} {\bibfnamefont {F.}~\bibnamefont {Lederer}},\ }\href@noop
  {} {\bibfield  {journal} {\bibinfo  {journal} {Physical review letters}\
  }\textbf {\bibinfo {volume} {96}},\ \bibinfo {pages} {053903} (\bibinfo
  {year} {2006})}\BibitemShut {NoStop}%
\bibitem [{\citenamefont {Shevchenko}\ \emph {et~al.}(2010)\citenamefont
  {Shevchenko}, \citenamefont {Ashhab},\ and\ \citenamefont
  {Nori}}]{shevchenko2010landau}%
  \BibitemOpen
  \bibfield  {author} {\bibinfo {author} {\bibfnamefont {S.}~\bibnamefont
  {Shevchenko}}, \bibinfo {author} {\bibfnamefont {S.}~\bibnamefont {Ashhab}},
  \ and\ \bibinfo {author} {\bibfnamefont {F.}~\bibnamefont {Nori}},\
  }\href@noop {} {\bibfield  {journal} {\bibinfo  {journal} {Physics Reports}\
  }\textbf {\bibinfo {volume} {492}},\ \bibinfo {pages} {1} (\bibinfo {year}
  {2010})}\BibitemShut {NoStop}%
\bibitem [{\citenamefont {Kayanuma}(1997)}]{kayanuma1997stokes}%
  \BibitemOpen
  \bibfield  {author} {\bibinfo {author} {\bibfnamefont {Y.}~\bibnamefont
  {Kayanuma}},\ }\href@noop {} {\bibfield  {journal} {\bibinfo  {journal}
  {Physical Review A}\ }\textbf {\bibinfo {volume} {55}},\ \bibinfo {pages}
  {R2495} (\bibinfo {year} {1997})}\BibitemShut {NoStop}%
\bibitem [{\citenamefont {Xiao}\ \emph {et~al.}(2010)\citenamefont {Xiao},
  \citenamefont {Chang},\ and\ \citenamefont {Niu}}]{xiao2010berry}%
  \BibitemOpen
  \bibfield  {author} {\bibinfo {author} {\bibfnamefont {D.}~\bibnamefont
  {Xiao}}, \bibinfo {author} {\bibfnamefont {M.-C.}\ \bibnamefont {Chang}}, \
  and\ \bibinfo {author} {\bibfnamefont {Q.}~\bibnamefont {Niu}},\ }\href@noop
  {} {\bibfield  {journal} {\bibinfo  {journal} {Reviews of modern physics}\
  }\textbf {\bibinfo {volume} {82}},\ \bibinfo {pages} {1959} (\bibinfo {year}
  {2010})}\BibitemShut {NoStop}%
\bibitem [{\citenamefont {Aidelsburger}\ \emph {et~al.}(2015)\citenamefont
  {Aidelsburger}, \citenamefont {Lohse}, \citenamefont {Schweizer},
  \citenamefont {Atala}, \citenamefont {Barreiro}, \citenamefont {Nascimbene},
  \citenamefont {Cooper}, \citenamefont {Bloch},\ and\ \citenamefont
  {Goldman}}]{aidelsburger2015measuring}%
  \BibitemOpen
  \bibfield  {author} {\bibinfo {author} {\bibfnamefont {M.}~\bibnamefont
  {Aidelsburger}}, \bibinfo {author} {\bibfnamefont {M.}~\bibnamefont {Lohse}},
  \bibinfo {author} {\bibfnamefont {C.}~\bibnamefont {Schweizer}}, \bibinfo
  {author} {\bibfnamefont {M.}~\bibnamefont {Atala}}, \bibinfo {author}
  {\bibfnamefont {J.~T.}\ \bibnamefont {Barreiro}}, \bibinfo {author}
  {\bibfnamefont {S.}~\bibnamefont {Nascimbene}}, \bibinfo {author}
  {\bibfnamefont {N.}~\bibnamefont {Cooper}}, \bibinfo {author} {\bibfnamefont
  {I.}~\bibnamefont {Bloch}}, \ and\ \bibinfo {author} {\bibfnamefont
  {N.}~\bibnamefont {Goldman}},\ }\href@noop {} {\bibfield  {journal} {\bibinfo
   {journal} {Nature Physics}\ }\textbf {\bibinfo {volume} {11}},\ \bibinfo
  {pages} {162} (\bibinfo {year} {2015})}\BibitemShut {NoStop}%
\bibitem [{\citenamefont {Drost}\ \emph {et~al.}(2017)\citenamefont {Drost},
  \citenamefont {Ojanen}, \citenamefont {Harju},\ and\ \citenamefont
  {Liljeroth}}]{Drost2017}%
  \BibitemOpen
  \bibfield  {author} {\bibinfo {author} {\bibfnamefont {R.}~\bibnamefont
  {Drost}}, \bibinfo {author} {\bibfnamefont {T.}~\bibnamefont {Ojanen}},
  \bibinfo {author} {\bibfnamefont {A.}~\bibnamefont {Harju}}, \ and\ \bibinfo
  {author} {\bibfnamefont {P.}~\bibnamefont {Liljeroth}},\ }\href@noop {}
  {\bibfield  {journal} {\bibinfo  {journal} {Nature Physics}\ }\textbf
  {\bibinfo {volume} {Advance online publication}},\ \bibinfo {pages} {4080}
  (\bibinfo {year} {2017})}\BibitemShut {NoStop}%
\bibitem [{\citenamefont {Slot}\ \emph {et~al.}(2017)\citenamefont {Slot},
  \citenamefont {Gardenier}, \citenamefont {Jacobse}, \citenamefont {van
  Miert}, \citenamefont {Kempkes}, \citenamefont {Zevenhuizen}, \citenamefont
  {Smith}, \citenamefont {Vanmaekelbergh},\ and\ \citenamefont
  {Swart}}]{slot2017experimental}%
  \BibitemOpen
  \bibfield  {author} {\bibinfo {author} {\bibfnamefont {M.~R.}\ \bibnamefont
  {Slot}}, \bibinfo {author} {\bibfnamefont {T.~S.}\ \bibnamefont {Gardenier}},
  \bibinfo {author} {\bibfnamefont {P.~H.}\ \bibnamefont {Jacobse}}, \bibinfo
  {author} {\bibfnamefont {G.~C.}\ \bibnamefont {van Miert}}, \bibinfo {author}
  {\bibfnamefont {S.~N.}\ \bibnamefont {Kempkes}}, \bibinfo {author}
  {\bibfnamefont {S.~J.}\ \bibnamefont {Zevenhuizen}}, \bibinfo {author}
  {\bibfnamefont {C.~M.}\ \bibnamefont {Smith}}, \bibinfo {author}
  {\bibfnamefont {D.}~\bibnamefont {Vanmaekelbergh}}, \ and\ \bibinfo {author}
  {\bibfnamefont {I.}~\bibnamefont {Swart}},\ }\href@noop {} {\bibfield
  {journal} {\bibinfo  {journal} {Nature Physics}\ }\textbf {\bibinfo {volume}
  {Advanced Online Publication}},\ \bibinfo {pages} {4105} (\bibinfo {year}
  {2017})}\BibitemShut {NoStop}%
\bibitem [{\citenamefont {Ozawa}\ \emph {et~al.}(2017)\citenamefont {Ozawa},
  \citenamefont {Taie}, \citenamefont {Ichinose},\ and\ \citenamefont
  {Takahashi}}]{ozawa2017interaction}%
  \BibitemOpen
  \bibfield  {author} {\bibinfo {author} {\bibfnamefont {H.}~\bibnamefont
  {Ozawa}}, \bibinfo {author} {\bibfnamefont {S.}~\bibnamefont {Taie}},
  \bibinfo {author} {\bibfnamefont {T.}~\bibnamefont {Ichinose}}, \ and\
  \bibinfo {author} {\bibfnamefont {Y.}~\bibnamefont {Takahashi}},\ }\href@noop
  {} {\bibfield  {journal} {\bibinfo  {journal} {Physical Review Letters}\
  }\textbf {\bibinfo {volume} {118}},\ \bibinfo {pages} {175301} (\bibinfo
  {year} {2017})}\BibitemShut {NoStop}%
\bibitem [{\citenamefont {Mur-Petit}\ and\ \citenamefont
  {Molina}(2014)}]{mur2014chiral}%
  \BibitemOpen
  \bibfield  {author} {\bibinfo {author} {\bibfnamefont {J.}~\bibnamefont
  {Mur-Petit}}\ and\ \bibinfo {author} {\bibfnamefont {R.~A.}\ \bibnamefont
  {Molina}},\ }\href@noop {} {\bibfield  {journal} {\bibinfo  {journal}
  {Physical Review B}\ }\textbf {\bibinfo {volume} {90}},\ \bibinfo {pages}
  {035434} (\bibinfo {year} {2014})}\BibitemShut {NoStop}%
\bibitem [{\citenamefont {Gong}\ \emph {et~al.}(2017)\citenamefont {Gong},
  \citenamefont {Higashikawa},\ and\ \citenamefont {Ueda}}]{gong2017zeno}%
  \BibitemOpen
  \bibfield  {author} {\bibinfo {author} {\bibfnamefont {Z.}~\bibnamefont
  {Gong}}, \bibinfo {author} {\bibfnamefont {S.}~\bibnamefont {Higashikawa}}, \
  and\ \bibinfo {author} {\bibfnamefont {M.}~\bibnamefont {Ueda}},\ }\href@noop
  {} {\bibfield  {journal} {\bibinfo  {journal} {Physical Review Letters}\
  }\textbf {\bibinfo {volume} {118}},\ \bibinfo {pages} {200401} (\bibinfo
  {year} {2017})}\BibitemShut {NoStop}%
\bibitem [{\citenamefont {Leykam}\ \emph {et~al.}(2012)\citenamefont {Leykam},
  \citenamefont {Bahat-Treidel},\ and\ \citenamefont
  {Desyatnikov}}]{leykam2012pseudospin}%
  \BibitemOpen
  \bibfield  {author} {\bibinfo {author} {\bibfnamefont {D.}~\bibnamefont
  {Leykam}}, \bibinfo {author} {\bibfnamefont {O.}~\bibnamefont
  {Bahat-Treidel}}, \ and\ \bibinfo {author} {\bibfnamefont {A.~S.}\
  \bibnamefont {Desyatnikov}},\ }\href@noop {} {\bibfield  {journal} {\bibinfo
  {journal} {Physical Review A}\ }\textbf {\bibinfo {volume} {86}},\ \bibinfo
  {pages} {031805} (\bibinfo {year} {2012})}\BibitemShut {NoStop}%
\bibitem [{\citenamefont {Aidelsburger}\ \emph {et~al.}(2012)\citenamefont
  {Aidelsburger}, \citenamefont {Atala}, \citenamefont {Barreiro},
  \citenamefont {Abanin}, \citenamefont {Kitagawa}, \citenamefont {Demler},\
  and\ \citenamefont {Bloch}}]{Aidelsburger2012Direct}%
  \BibitemOpen
  \bibfield  {author} {\bibinfo {author} {\bibfnamefont {M.}~\bibnamefont
  {Aidelsburger}}, \bibinfo {author} {\bibfnamefont {M.}~\bibnamefont {Atala}},
  \bibinfo {author} {\bibfnamefont {J.~T.}\ \bibnamefont {Barreiro}}, \bibinfo
  {author} {\bibfnamefont {D.}~\bibnamefont {Abanin}}, \bibinfo {author}
  {\bibfnamefont {T.}~\bibnamefont {Kitagawa}}, \bibinfo {author}
  {\bibfnamefont {E.}~\bibnamefont {Demler}}, \ and\ \bibinfo {author}
  {\bibfnamefont {I.}~\bibnamefont {Bloch}},\ }\href@noop {} {\bibfield
  {journal} {\bibinfo  {journal} {Nature Physics}\ }\textbf {\bibinfo {volume}
  {9}},\ \bibinfo {pages} {795} (\bibinfo {year} {2012})}\BibitemShut {NoStop}%
\bibitem [{\citenamefont {Szameit}\ \emph {et~al.}(2006)\citenamefont
  {Szameit}, \citenamefont {Burghoff}, \citenamefont {Pertsch}, \citenamefont
  {Nolte}, \citenamefont {T{\"u}nnermann},\ and\ \citenamefont
  {Lederer}}]{szameit2006two}%
  \BibitemOpen
  \bibfield  {author} {\bibinfo {author} {\bibfnamefont {A.}~\bibnamefont
  {Szameit}}, \bibinfo {author} {\bibfnamefont {J.}~\bibnamefont {Burghoff}},
  \bibinfo {author} {\bibfnamefont {T.}~\bibnamefont {Pertsch}}, \bibinfo
  {author} {\bibfnamefont {S.}~\bibnamefont {Nolte}}, \bibinfo {author}
  {\bibfnamefont {A.}~\bibnamefont {T{\"u}nnermann}}, \ and\ \bibinfo {author}
  {\bibfnamefont {F.}~\bibnamefont {Lederer}},\ }\href@noop {} {\bibfield
  {journal} {\bibinfo  {journal} {Optics express}\ }\textbf {\bibinfo {volume}
  {14}},\ \bibinfo {pages} {6055} (\bibinfo {year} {2006})}\BibitemShut
  {NoStop}%
\end{thebibliography}%

\clearpage
\newpage
 \vskip0.25cm\noindent
{\bf Acknowledgements}\\
This work is supported by the National Youth 1000 Talents Program in China, and the 985 startup Grant (205020516074) at Tongji University.

 \vskip0.25cm\noindent
{\bf Author contributions}\\
J.R. planned and guided the research. Y. L. performed the calculating and drawing. Both authors contributed to the discussions and the manuscript. 

 \vskip0.25cm\noindent
{\bf Competing financial interests:} The author declares no competing financial interests.


\begin{widetext}

\section{ Supplemental Material - Topological Landau-Zener Bloch Oscillations in Photonic Floquet Lieb Lattices}

\section{TIGHT-BINDING MODEL}
The dynamics of the field on this lattice can be described by schr$\ddot{o}$dinger equation:
\begin{equation}
i \frac{d}{dz} |\psi\rangle = \hat{H}(z)|\psi \rangle
\end{equation}
where $|\psi\rangle$ is the photon wave function. By using fourier series, we can transform real-space tight binding Hamiltonian results in a $k$-space Hamiltonian which is $3*3$ matrix:
\begin{equation}
\hat{H}(z) = \sum_{\bm{k}} \begin{pmatrix} \hat{a}^{\dagger}_{\bm{k}}\ \hat{b}^{\dagger}_{\bm{k}}\  \hat{c}^{\dagger}_{\bm{k}} \end{pmatrix} H(\bm{k},\bm{A}(z)) \begin{pmatrix}
\hat{a}_{\bm{k}} \\
\hat{b}_{\bm{k}} \\
\hat{c}_{\bm{k}} \\
\end{pmatrix}
\end{equation}
and
\begin{equation}
H(\bm{k},\bm{A}(z)) = \begin{pmatrix}
     0 & H_1(z) & 0\\
     H_1^{*}(z) & 0 & H_2(z)\\
     0 & H_2^{*}(z) & 0
\end{pmatrix}
\label{eq:liefloquetbHamiltonian}
\end{equation}
where $H_1(z) = \kappa_1 exp(i(k_x+A_x))+\kappa_2 exp(-i(k_x+A_x))$, $H_2(z) = \kappa_1 exp(i(k_y+A_y))+ \kappa_2 exp(-i(k_y+A_y))$, $k_x$ and $k_y$ is the component of Bloch wavevecter $\bm{k}$, and the hopping energy $\kappa_1 \ne \kappa_2$. Obviously, in this case, $H(z+Z_0)=H(z)$, thus the solution of Eq.\ref{eq:liefloquetbHamiltonian} can be written as $|\psi (z)\rangle = e^{-i\beta z}|\chi (z)\rangle$, where $|\chi(z+2\pi/\Omega)\rangle = |\chi(z)\rangle$, where $\epsilon (mod\quad\Omega)$ is the quasi-energy. We can expand the $|\psi (z)\rangle$ as following fourier series form:
\begin{equation*}
|\chi (z)\rangle = \sum_{n=-\infty}^{\infty} |\chi_n\rangle e^{in\Omega t}
\end{equation*}
where $\Omega = \frac{Z_0}{2\pi}$ and $|\chi_n\rangle$ is the fourier coefficient of $|\chi (z)\rangle$. Considering least order form and ignore the higher order ($|n|\geq 2$), we can expand the periodic-driven gauge field into the form:
\begin{eqnarray*}
& e^{\pm i A_0 sin(\omega z)} \approx J_0(A_0) \mp J_1(A_0) e^{i\phi} e^{-i \omega z} \mp J_1(A_0) e^{-i\phi}  e^{i \omega z}\\
& e^{\pm i A_0 sin(\omega z-\phi)} \approx J_0(A_0) \mp J_1(A_0) e^{i\phi}  e^{-i \omega z} \mp J_1(A_0) e^{-i\phi}  e^{i \omega z}\\
\end{eqnarray*}
where the function $J_0$ and $J_1$ are the Bessel functions of first kind with the order 0 and 1. The components of effective $k$-space Hamiltonian can be written as:
\begin{widetext}
\begin{equation*}
\begin{split}
 H_0 =  
 J_0(A_0) \begin{pmatrix}
     0 & \kappa_1 e^{i k_x}+\kappa_2 e^{-i k_x} & 0\\
     \kappa_1 e^{-i k_x}+\kappa_2 e^{i k_x} & 0 & \kappa_1 e^{i k_y} + \kappa_2 e^{-i k_y} \\
     0 & \kappa_1 e^{-i k_y} + \kappa_2 e^{i k_y} & 0
\end{pmatrix} \\
H_1 = J_1(A_0)
\begin{pmatrix}
     0 & -\kappa_1 e^{i k_x}+\kappa_2 e^{-i k_x} & 0\\
     \kappa_1 e^{-i k_x}-\kappa_2 e^{i k_x} & 0 & e^{i \phi}(-\kappa_1 e^{i k_y} + \kappa_2 e^{-i k_y}) \\
     0 & e^{i \phi}(\kappa_1 e^{-i k_y} - \kappa_2 e^{i k_y}) & 0
	\end{pmatrix} \\
H_{-1} = J_1(A_0)
\begin{pmatrix}
     0 & \kappa_1 e^{i k_x}-\kappa_2 e^{-i k_x} & 0\\
     -\kappa_1 e^{-i k_x}+\kappa_2 e^{i k_x} & 0 & e^{-i \phi}(\kappa_1 e^{i k_y} - \kappa_2 e^{-i k_y}) \\
     0 & e^{-i \phi}(-\kappa_1 e^{-i k_y} + \kappa_2 e^{i k_y}) & 0
	\end{pmatrix}
\end{split}
\end{equation*}
\end{widetext}

By exploiting the formula\cite{cayssol2013floquet} $H_{eff} = H_0 + \frac{1}{\omega} [H_1,H_{-1}] $, we can get the effective Hamiltonian for the periodic-driven Floquet Lieb lattice system as:
\begin{widetext}
\begin{equation}
H_{eff} = \Delta_0 \begin{pmatrix}
0 & \kappa_1 e^{i k_x} + \kappa_2 e^{-i k_x} & i\tau (\kappa_1 e^{i k_x} - \kappa_2 e^{- i k_x})(\kappa_1 e^{i k_y} - \kappa_2 e^{-i k_y}) \\
\kappa_1 e^{-i k_x} + \kappa_2 e^{i k_x}  & 0 & \kappa_1 e^{i k_y} + \kappa_2 e^{-i k_y} \\
- i\tau (\kappa_1 e^{-i k_x} - \kappa_2 e^{ i k_x})(\kappa_1 e^{-i k_y} - \kappa_2 e^{i k_y}) & \kappa_1 e^{-i k_y} + \kappa_2 e^{i k_y} & 0
\end{pmatrix}
\label{eq:effectivehamltonian}
\end{equation}
\end{widetext}
where $\Delta_0 = J_0(A_0)$, $\Delta_1 = -\frac{J_1(A_0)^2 sin(\phi)}{2 \omega}$, and $\tau=\frac{\Delta_1}{\Delta_0}$.

\begin{figure}[htbp]
\centering
\includegraphics[width=\linewidth]{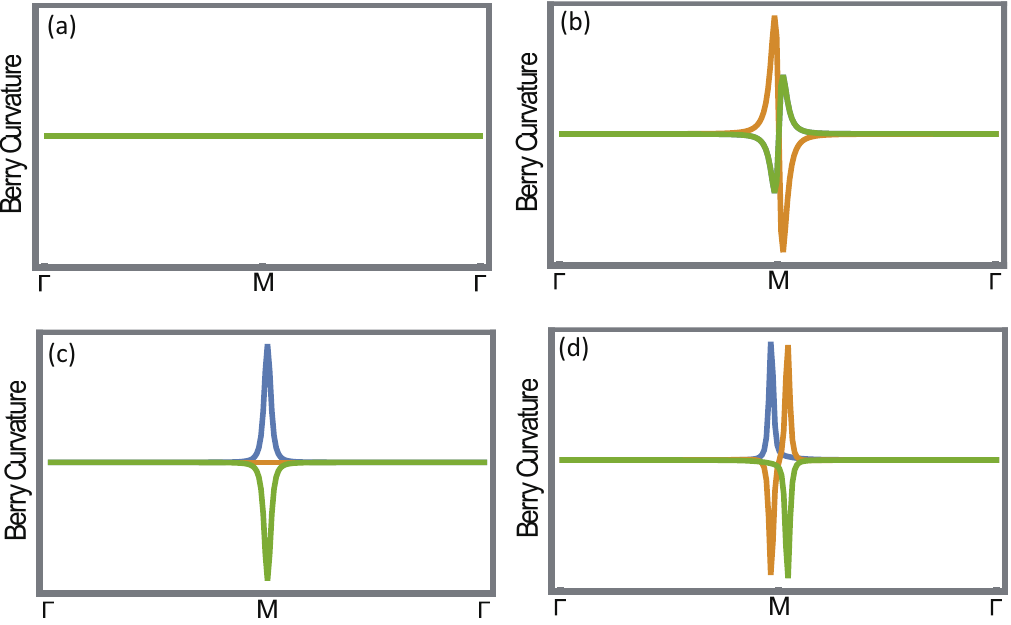}
\caption{Berry Curvature distribution in momentum space. (a) Standard lieb lattice of waveguides. ($\kappa_1=\kappa_2$, $A_0=0$). (b) Dimertized lieb lattice of waveguides. ($\kappa_1\ne \kappa_2$, $A_0=0$). (c) Standard lieb lattice of helical waveguides. ($\kappa_1= \kappa_2$, $A_0\ne0$). (d) Dimertized lieb lattice of helical waveguides. ($\kappa_1\ne \kappa_2$, $A_0\ne0$). The upper band in band structure is plotted in blue color, flat band in orange color, and lower band in green color, respectively.}
\label{fig:berrycurvature}
\end{figure}

When $\kappa_1 = \kappa_2$, the periodic-driven guage field will induce topological non-trivial gaps in optical Lieb lattice system. The $k$-space effective Hamiltonian Eq.(\ref{eq:effectivehamltonian}) around the M point in  Brillouin zone can be written as the form:
\begin{equation*}
\begin{split}
H_{eff} = 2 J_0(A_0) \kappa_1  k_x S_x + 2 J_0(A_0) \kappa_1 k_y S_y \\
 +\frac{2 J_1(A_0)^2 \kappa_1^2 sin(\phi)}{\omega} S_z
\end{split}
\end{equation*}
where $\bm{k}=(k_x,k_y)$ is the Bloch wave vector and $\bm{S} = (S_x,S_y,S_z)$ are pseudospin operators satisfying the angular momentum algebra $[S_i,S_j] = i \varepsilon_{ijk} S_k$.  
\begin{equation*}
S_x = \begin{pmatrix}
0 & 1 & 0 \\
1 & 0 & 0 \\
0 & 0 & 0
\end{pmatrix},
S_y = \begin{pmatrix}
0 & 0 & 0 \\
0 & 0 & 1 \\
0 & 1 & 0
\end{pmatrix},
S_z = \begin{pmatrix}
0 & 0 & i \\
0 & 0 & 0 \\
-i & 0 & 0
\end{pmatrix}
\end{equation*}
When $\phi \ne 0$, the topological non-trivial gaps will open at the point $M$. The sizes of these gaps are determined by the strength $A_0$ and modulation frequency $\Omega$ of periodic-driven gauge field. Specially, the flat bands maintain flat even with Floquet mechanism and its Berry curvature keeps zero in whole Brillouin zone shown as Fig.\ref{fig:berrycurvature}(c).  Distinguished with the spin-$\frac{1}{2}$ system, i.e, Dirac fermions in graphene lattice, the composite Lieb bosons\cite{leykam2012pseudospin,diebel2016conical} supported by the Lieb lattice are the spin-1 system and the spin-0 states always stay constant energies even with non-trivial topological transition. But it should be mentioned that for flat band it will be connect with other bands and can not be isolated after topological transition.

When $\kappa_1 \ne \kappa_2$, there exist topological trivial gaps without the effective gauge field $\bm{A}(z)$ due to the dimertized nearest-neigbhor coupling. In this case, the coupling setting is similar with the SSH model in two dimension, but according to the topological periodic table, 2D SSH model belongs to the type BDI and no topological transition can be induced after introducing the dimertized couplings. But with effective gauge field, i.e, $A_0 \ne 0$, the system will be different. The original topological band gap will be close and reopen with the influence of effective gauge field increasing. Different with equality case($\kappa_1 = \kappa_2$), the flat will become perturbed yet gapless structure with almost flat parts. The reason behind this is that the dimertized nearest-neighbor coupling $\kappa_1 \ne \kappa_2$ leads to lower spectral symmetry and non-zero effective gauge field $\bm{A}$ induces time-reversal broken non-trivial geometric phase. 

By comparing the equality case $\kappa_1 = \kappa_2$ and dimertized case $\kappa_1 \ne \kappa_2$, we can find that the topological transitions of the equality case for three bands happen at the same point $M$. Because the local symmetries of unit cell make sure that two topological transition processes for two gaps can't be separated and recognized by inducing effective gauge field. But for the dimertized case, the situation will be different. Lower of the local symmetries of unit cell will make the two processes separated and thus two kinds of topological non-trivial gaps open in two different points in momentum space. This difference can be viewed from the Berry curvature distribution for flat bands in Fig.\ref{fig:berrycurvature}(c)(d), for equality case, the Berry curvature keeps zero in whole Brillouin zone while it will have a time-reversal broken distribution for dimertized case.

It's important to mention that the topological non-trivial gap sizes can be modulated by the phase $\phi$. The energy shift for each spin is proportional to $sin(\phi)$, which means positive shift happens for $\phi \in (0,\pi)$ and negative happens for $\phi \in (0,-\pi)$. For the special case $\phi = 0, \pi$, the system keeps same even with non-zero effective gauge field strength $A_0 \ne 0$, but the effects accumulated from the Floquet mechanism vanish.

\subsection{Landau-Zener Tuneling}
To study the non-adiabatic process during Bloch oscillation, we consider static case that $A_0=0$,  the Floquet effective Hamiltonian $H_{eff}$ will represent the basic dynamic coupling of the basic modes.  The corresponding effective Hamiltonian without Floquet mechanism is:
\begin{equation*}
H_{s} = \begin{pmatrix}
0 & \alpha & 0 \\
\alpha*  & 0 & \beta \\
0 & \beta* & 0
\end{pmatrix}
\end{equation*}
where $\alpha = \kappa_1 e^{i k_x} + \kappa_2 e^{-i k_x}$ and $\beta = \kappa_1 e^{i k_y} + \kappa_2 e^{-i k_y}$. Based on this Hamiltonian, we can calculate the corresponding eigenvectors $|S_{+}\rangle$, $|S_{0}\rangle$, $|S_{-}\rangle$. By multiplying $|\psi\rangle$ with the conjugate transpose of the matrix $U= [|S_{+}\rangle,|S_{0}\rangle,|S_{-}\rangle]$, and
\begin{equation*}
H' = U^{-1} H_{eff} U
\end{equation*}
The floquet effective Hamiltonian can be expressed in the new basis of $|S_{+}\rangle$, $|S_{0}\rangle$, $|S_{-}\rangle$ and the wave function can be represented by the linear combination of new basis $|\psi\rangle = c_{+}|S_{+}\rangle + c_{0} |S_{0}\rangle + c_{-}|S_{-}\rangle$, where $c_{+}$,$c_{0}$,$c_{-}$ are complex number.

With effective external force field $\bm{F} = (F_x,F_y)$, we set $k_x = F_x t$, $k_y = F_y t$.So we have dynamic equation:
\begin{equation}
i \frac{\partial}{\partial z} \begin{pmatrix}
c_{+} \\
c_{0} \\
c_{-}
\end{pmatrix}
=  H'
\begin{pmatrix}
c_{+} \\
c_{0} \\
c_{-}
\end{pmatrix}
\label{eq:landauzenerdynamic}
\end{equation}
Landau-Zener tuneling will take place at the avoided band crossings around $M$ point. Expanding Eq.\ref{eq:landauzenerdynamic} around $M$ point, we can get the triple Zener tuneling equations, and under the condition $\kappa_1 \ne \kappa_2$,  Landau-Zener tunneling takes place at slightly shifted positions for upper and lower bands in Fig.(2)(d), which reduces the problem into two standard Landau-Zener models:
\begin{eqnarray}
i\frac{\partial}{\partial z}c_{+} = - \frac{\epsilon_0 z}{2}c_{+} + \Delta\epsilon c_{0},\quad i\frac{\partial}{\partial z}c_{0} = \frac{\epsilon_0 z}{2} c_{0} + \Delta\epsilon c_{+} \\
i\frac{\partial}{\partial z}c_{0} = - \frac{\epsilon_0 z}{2}c_{0} + \Delta\epsilon c_{-},\quad i\frac{\partial}{\partial z}c_{-} = \frac{\epsilon_0 z}{2} c_{-} + \Delta\epsilon c_{0} 
\end{eqnarray}
where $\epsilon_0 = \sqrt{2}|\kappa_1-\kappa_2|$ and $2\Delta\epsilon$ is the gap between the flat and the dispersive bands. The tunneling probability is calculated according to Landau-Zener formula: $P_{LZ} = exp(-2\pi\Delta\epsilon^2/\epsilon_0)$. When considering the interference involved in this non-adiabatic process, on can keep track of this phase by describing the Landau-Zener transition using the non-adiabatic unitary evolution matrix $N$:
\begin{equation*}
\begin{pmatrix}
c_{+}(z_0+0) \\
c_{0}(z_0+0)
\end{pmatrix}
=  N
\begin{pmatrix}
c_{+}(z_0-0) \\
c_{0}(z_0-0)
\end{pmatrix}
\end{equation*}
\begin{equation*}
N = \begin{pmatrix}
\sqrt{1-P_{LZ}}e^{-i \varphi_{S}} & -\sqrt{P_{LZ}}\\
\sqrt{P_{LZ}} & \sqrt{1-P_{LZ}}e^{i \varphi_{S}}
\end{pmatrix}
\end{equation*}
\begin{equation}
\varphi_{S} = - \frac{\pi}{4} + \delta(ln\delta - 1) + arg\Gamma(1-i\delta)
\end{equation}
where $\varphi_{S}$ is called Stokes phase\cite{kayanuma1997stokes}, $\Gamma$ is the Gamma function and $\delta = \frac{\Delta\epsilon^2}{\epsilon_0}$. Same form for the case $c_{0}$ and $c_{-}$.

\subsection{Constructive and destructive interference}
To consider the all effect raise from the Bloch oscillation, we consider the whole floquet Lieb lattice subject subject to the effective external force field $\bm{F}$, which can be described by the following Hamiltonian:
\begin{widetext}
\begin{eqnarray*}
H_F = H_{eff}(\bm{k}) + \sum_{<n,m>} \{((nd-x_0)F_x + ((m+1/2)d-y_0)F_y)a_{n,m}^{\dagger} a_{n,m} + ((nd-x_0)F_x+(md-y_0)F_y)b_{n,m}^{\dagger}b_{n,m} \\
+ (((n+1/2)d-x_0)F_x+(md-y_0)F_y)c_{n,m}^{\dagger}c_{n,m} \}
\end{eqnarray*}
\end{widetext}
Eigenstates of $H_{eff}(\bm{k})$ can be expressed as the form $|u_{l,\bm{k}}\rangle = [\alpha_{l,\bm{k}}, \beta_{l,\bm{k}},\gamma_{l,\bm{k}}]^{T}$ with energy $\beta_{l,\bm{k}}$, where $l=1,2,3$ indicates the three bands. For discussing the difference of Bloch oscillation before and after topological transition, it is convenient to work in second quantization:

\begin{equation*}
\begin{split}
\varphi_{l,\bm{k}}^{\dagger} = \frac{1}{\sqrt{MN}}\sum_{<n,m>} \{+ \alpha_{l,\bm{k}} e^{i((n+\frac{1}{2})k_xd + mk_yd)}a_n^{\dagger} \\
+ \beta_{l,\bm{k}} e^{i(nk_xd+mk_yd)b_n^{\dagger}} + \gamma_{l,\bm{k}} e^{i(nk_xd+(m+\frac{1}{2})k_yd)} c_n^{\dagger}\}
\end{split}
\end{equation*}
where $M$,$N$ are the total number of sites in lattice along $x$ and $y$, respectively. For finite effective external force field $\bm{F}$, we can solve the Heisenberg-like equation of motion $i\frac{d}{dz}\Psi^{\dagger}(z) = [H_F,\Psi^{\dagger}(z)]$. We look for the solutions where the quasi-momentum changes at a constant rate, $k_x = k_{x,0} + F_x z$ and $k_y = k_{y,0} + F_y z$,
\begin{equation*}
\Psi^{\dagger}(z) = A_1(z)\varphi_{1,\bm{k}}^{\dagger} + A_2(z)\varphi_{2,\bm{k}}^{\dagger} + A_3(z)\varphi_{3,\bm{k}}^{\dagger}
\end{equation*}

Based on the perturbation approximation, for simplicity, only consider the case $ F = F_x = F_y$, which means the direction of force field $\bm{F}$ is along $M$ point, we can get a solution of the Heisenberg-like equation of motion\cite{Aidelsburger2012Direct}:
\begin{widetext}
\begin{eqnarray}
\label{eq:heisenbergequation1}
i\frac{\partial}{\partial z}A_1 = \beta_{1,\bm{k}}A_1 + Fx_0A_1 + iF \langle u_{1,\bm{k}}|\partial_{\bm{k}} u_{1,\bm{k}}\rangle + iF\langle u_{1,\bm{k}}|\partial_{\bm{k}} u_{2,\bm{k}}\rangle + iF\langle u_{1,\bm{k}}|\partial_{\bm{k}} u_{3,\bm{k}}\rangle \\
\label{eq:heisenbergequation2}
i\frac{\partial}{\partial z}A_2 = \beta_{2,\bm{k}}A_2 + Fx_0A_2 + iF \langle u_{2,\bm{k}}|\partial_{\bm{k}} u_{1,\bm{k}}\rangle + iF\langle u_{2,\bm{k}}|\partial_{\bm{k}} u_{2,\bm{k}}\rangle + iF\langle u_{2,\bm{k}}|\partial_{\bm{k}} u_{3,\bm{k}}\rangle \\ 
\label{eq:heisenbergequation3}
i\frac{\partial}{\partial z}A_3 = \beta_{3,\bm{k}}A_3 + Fx_0A_3 + iF \langle u_{3,\bm{k}}|\partial_{\bm{k}} u_{1,\bm{k}}\rangle + iF\langle u_{3,\bm{k}}|\partial_{\bm{k}} u_{2,\bm{k}}\rangle + iF\langle u_{3,\bm{k}}|\partial_{\bm{k}} u_{3,\bm{k}}\rangle
\end{eqnarray}
\end{widetext}
According to the Eq.\ref{eq:heisenbergequation1}, the first term describes the dynamical phase contribution, the second comes from the external effective force field contribution, effectively, Zeeman phase, the third term is the Berry's phase part, and the last term represents the phase contributed by non-adiabatic process.

When the light is evolved through the Brillouin zone of the periodic potential, $\bm{k} \rightarrow \bm{k}+\bm{G}$, it acquires a phase shift due to three distinct distributions: a geometric phase $\varphi_{g}$, the St\"uckelberg phase $\varphi_{St}$\cite{shevchenko2010landau} and a phase $\varphi_{F}$ due to energy bias of the sites in the external effective force field(which can be vanished by setting appropriate position $x_0$ and $y_0$).
\begin{equation*}
\varphi_{tot} = \varphi_{g} + \varphi_{St} + \varphi_{F}
\end{equation*}
where the St\"uckelberg phase $\varphi_{St}$ consists of two components: the dynamic phase $\varphi_{dyn}$ acquired the adiabatic evolution and the stokes phase $\varphi_{S}$ during the non-adiabatic evolution. 

Before the topological transition, the components $\langle u_{1,\bm{k}}|\partial u_{1,\bm{k}}\rangle$, $\langle u_{2,\bm{k}}|\partial u_{2,\bm{k}}\rangle$ and $\langle u_{3,\bm{k}}|\partial u_{3,\bm{k}}\rangle$ in the Eqs.(\ref{eq:heisenbergequation1},\ref{eq:heisenbergequation2},\ref{eq:heisenbergequation3}) is trivial, the geometric phase $\varphi_{g}$ accumulated in the Bloch oscillation is zero. The phase accumulated during the Bloch oscillation can only reflect the basic dynamic and tuneling properties. The interference between different passages is only based on the St\"uckelberg phase $\varphi_{St}$\cite{kling2010atomic,shevchenko2010landau}.While the bulk bands close and reopen topological non-trivial gaps, the components $\langle u_{1,\bm{k}}|\partial u_{1,\bm{k}}\rangle$, $\langle u_{2,\bm{k}}|\partial u_{2,\bm{k}}\rangle$ and $\langle u_{3,\bm{k}}|\partial u_{3,\bm{k}}\rangle$ in the Eqs.(\ref{eq:heisenbergequation1},\ref{eq:heisenbergequation2},\ref{eq:heisenbergequation3}) can't vanish. The geometric phase $\varphi_{g}$ accumulated in the Bloch oscillation for each band is associated with its Chern number, dependent with the distribution of Berry curvature in momentum space(mostly concentrate on the vicinity region of $M$ point for our model). The geometric phase $\varphi_{g}$ will involve in interfering with the St\"uckelberg phase $\varphi_{St}$. The Bloch oscillation in non-trivial case will be in different interference situation.

\section{CONTINUUM MODEL}
Similar to real experiments, the continuum model uses same parameters with Ref.\cite{rechtsman2013photonic}, describing femtosecond laser-written waveguide arrays in fused silica. The optical behavior can be described by the beam propagation method,
\begin{equation}
i \frac{\partial \psi}{\partial z} = - \frac {1}{2n_0k_0} \nabla^2 \psi - k_0 \Delta n(x,y,z) \psi
\end{equation}
where $k_0 = 2\pi/\lambda$, the background refractive index is $n_0=1.45$ at wavelength $\lambda = 633nm$. Same descriptions with Ref.\cite{leykam2016anomalous, rechtsman2013photonic}, the index modulation forms a square helix lattice:
\begin{equation}
\begin{split}
\Delta n (x,y,z) = \Delta n_0 \sum_{m,n} \{ V_0(X_{n+1/4},Y_{n+1/4}) \\
+V_0(X_{n+1/4},Y_{n+3/4}) \\
+V_0(X_{n+3/4},Y_{n+1/4}) \}
\end{split}
\end{equation}
where
\begin{equation}
\begin{split}
X_n(z) = x-x_0-n a \\
Y_m(z) = y-y_0-m a
\end{split}
\end{equation}
$\Delta n_0 = 7*10^{-4}$ is the modulation index, $a$ is the lattice constant, and $V_0(x,y) = exp(-((2x/L_x)^2+(2y/L_y)^2)^3)$ with width $ L_x = 11\mu m$ and $L_y = 4\mu m$.  According to real experiments, the refractive index is $n_0 = 1.45$ at wavelength $\lambda = 633 nm$, with modulation $\Delta n = 7.0 \times 10^{-4}$. The waveguides elliptical cross sections with axis diameters 11 and 4 $\mu m$, as shown in Fig.1(b) of main text. To induce the appropriate gauge field for photon, we can adjust  the lattice constant $a$, helix radius $R$, and pitch $Z_0$ of Floquet Lieb lattice. The distant between nearest waveguides along $x$ and $y$ direction, i.e, $d_1 \ne d_2$ will be different can be used to achieve dimertized nearest-neigbhor coupling $\kappa_1 \ne \kappa_2$. In our model, we set $a = 50\mu m$, $R = 3 \mu m$, $Z_0 = 1mm $, $d_1 = (1/2 + 1/40)a$ and $d_2 = (1/2 - 1/40)a$.

\subsection{Floquet Band Structure Calculation}
Due to the quasienergies $\beta$ are defined modulo $2\pi/Z_0$ and hard to converge on during computation, direct calculation of the Floquet band  structure for a continuum model (proposed by tight-binding model) is a non-trivial task. Here, we perform the calculation by calculating the variation of evolution operator $U(\bm{k},z)$ with the axis $z$ and finally get the Bloch mode evolution operator $\hat{U}(\bm{k},Z) = e^{-i \int_0^Z \hat{H}(\bm{k},z) dz }$, $|u_n(\bm{k},Z)\rangle = \hat{U}(\bm{k},Z)|u_n(\bm{k},0)\rangle$.

Here we split the unit cell into $M$ pieces $\{z_i\}$ along $z$ direction, $i = 1, ..., M$ and calculate the their quasi-eigenenergy, respectively. We obtain the eigen-modes by solving the static Bloch mode eigenvalue problem at $z_i$,
\begin{equation}
E_n(\bm{k},z_i)|\psi_n(\bm{k},z_i) \rangle = \hat{H}(\bm{k},z_i)|\psi_n(\bm{k},z_i) \rangle
\end{equation}
where $\hat{H}(\bm{k},z_i)=-\frac{1}{2n_0k_0}\nabla^2 - k_0 \Delta n(x,y,z_i)$ is the Bloch Hamiltonian and $\bm{k}$ is the Bloch momentum, $\bm{k}=(k_x,k_y)$. For waveguides fabricated by the femtosecond laser writing technique, it can be chosen to be single lowest mode, with good confinement. Thus, the number of bands $N$ can be limited as the number of waveguides per unit cell ($N=3$). Numerically, we can calculate the truncated basis ${|\psi_n(\bm{k},z_i\rangle) }$, $n=1,...,N$ at any $z_i$ in unit cell. With Floquet mechanism, the system is periodic for one modulation period, which requires:

\begin{equation}
\begin{split}
|\psi_n(\bm{k}, Z_0) \rangle = e^{-i \int_0^{Z_0} \hat{H}(\bm{k},z)} dz |\psi_n(\bm{k},0) \rangle \\
= e^{-i H_{eff} Z_0} |\psi_n(\bm{k},0)\rangle \\
= \hat{U}(\bm{k},Z_0) |\psi_n(\bm{k},0\rangle
\end{split}
\end{equation}
where $\hat{U}(\bm{k},Z_0)$ is the Bloch mode evolution operator. For any $z_i$ in unit cell, $\sum_{n=1}^{N} |\psi_n(\bm{k},z_i)\rangle\langle\psi_n(\bm{k},z_i)| = 1$. We can obtain evolution operator by expanding it with truncated evolution operator at $z_i$ and $z_i+dz$ with small step $dz$ :
\begin{equation}
U_{mn}(\bm{k},z_i) = \langle \psi_m(\bm{k},z_i+dz)|\hat{U}(\bm{k},z_i)|\psi_n(\bm{k},z_i)\rangle
\end{equation}
Here we assumes that the evolution $\hat{U}$ does not couple modes beyond the $N$th band. After calculating the $U_{mn}(\bm{k},z_i)$, we can get the formula of $\hat{U}(\bm{k},Z_0)$ directly:
\begin{equation}
\hat{U}(\bm{k},Z_0) = \prod_{i=1}^{M} U(\bm{k},z_i) E(\bm{k},z_i)
\end{equation}
where $E(\bm{k},z_i) = diag(E_1(\bm{k},z_i), E_2(\bm{k},z_i), ... ,E_N(\bm{k},z_i))$. Given $\hat{U}(\bm{k},Z_0)$, we can compute its eigenvectors, berry phase and hence evaluate the Chern number numerically. Meanwhile, the effective Hamiltonian can be determined by $H_{eff}=\frac{i}{Z_0} log(\hat{U}(\bm{k},Z_0))$. 

To fit the parameters of the tight binding band structure to the continuum model's in Fig.(2) of main text, we set $\kappa_1, \kappa_2, A_0$ to adjust the bandstructure especially at point $M$ and besides these basic parameters, we also need to set appropriate on-site energy to fit the energy level. This is already sufficient to demonstrate a reasonable agreement between the two models. Another intuitive and efficient way to obtain a correspondence between the continuum and TB model parameters is to exploit a simplified case, i.e, a single two-waveguide coupler to calculate its optical behavior and then extract $\kappa_1, \kappa_2, A_0$. Therefore, the optimization of two-waveguide coupler will optimize the design parameters of Floquet Lieb lattice(eg. lattice constant, separation). 

\subsection{Effective external force field}
The key point to achieve Bloch oscillation is to realize effective external force field in optical system. The convenient way is to change the refraction index of sites in sublattice, which can be realized by adjusting the power of femtosecond laser when writing the waveguide arrays in fused silica\cite{szameit2006two}. The variation of refractive index can be represented as\cite{trompeter2006bloch}
\begin{equation*}
\Delta n(x,y) = \gamma \frac{I(x,y)}{1+I(x,y)}
\end{equation*}
where  $\gamma$ is constant number, and $I(x,y)$ is the position-dependent laser power. To realized effective force field, we can set the index distribution to satisfy the following condition:
\begin{equation*}
\frac{\Delta n(x,y) - \Delta n_{min}}{\Delta n_{max}- \Delta n_{min}} =  \frac{(x-x_0)F_x}{\Theta_x} + \frac{(y-y_0)F_y}{\Theta_y} 
\end{equation*}
where $\Theta_x$ and $\Theta_y$ is the ratio constant to control the index gradient.

\section{Unidirectional EDGE STATES}
After topological transition, the model system will transfer into a Floquet topological insulator, which exists one-way spin-polarized edge mode. In this case, we can calculate the band structure for edge state by using the following Hamiltonian:
\begin{equation}
\begin{split}
H(k_y, z) = \sum_{\langle n \rangle} (\kappa_y e^{i (A_y + k_y)} b^{\dagger}_{n} a_{n} + \kappa_y e^{i(k_y-A_y)} b^{\dagger}_{n} a_{n}  \\ + \kappa_x e^{iA_x} c^{\dagger}_{n} a_{n} + \kappa_x e^{-iA_x} c^{\dagger}_{n-1} a_{n}) + h.c.
\label{eq:liebedgeHamiltonian}
\end{split}
\end{equation}
Along $y$ direction, the model is periodic and therefore $k_y$ will be a good quantum number to describe the system. The edge state can be represented by the wave function: $|\phi(k_y)\rangle = \{a^{\dagger}_1a_1,b^{\dagger}_1b_1,c^{\dagger}_1c_1, ...,a^{\dagger}_Na_N,b^{\dagger}_Nb_N,c^{\dagger}_Nc_N\}^T$. 

\begin{figure}[htbp]
\centering
\includegraphics[width=\linewidth]{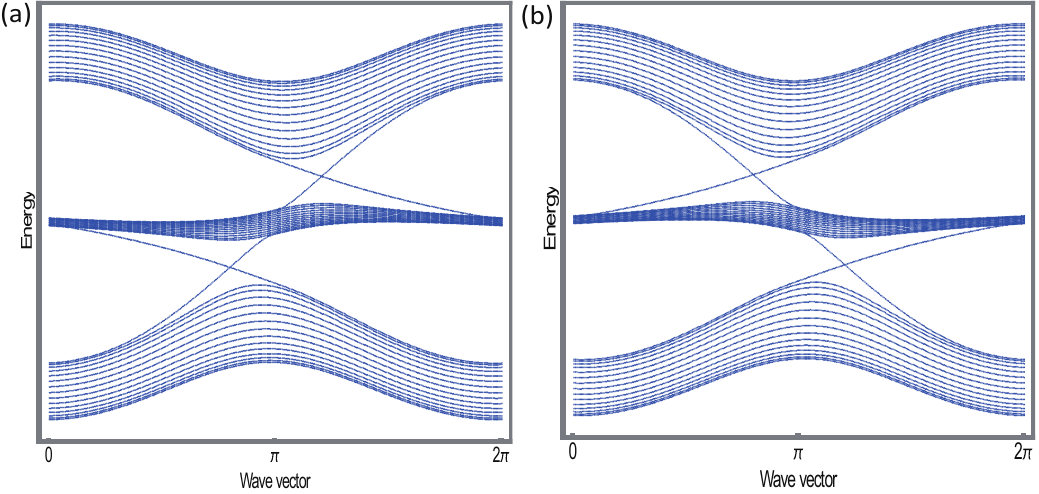}
\caption{Band structure for a semi-infinite strip ten unit cells wide. Blue circle points are obtained from the continuum model, red curves from the tight-binding model.(a)For the case $\phi=\pi/2$. (b)For the case $\phi=-\pi/2$.}
\label{fig:edgebands}
\end{figure}

The band structure for a semi-infinite strip ten unit cells wide have been shown in Fig.(\ref{fig:edgebands}). For the case that Dimertized lieb lattice of helical waveguides ($\kappa_1\ne \kappa_2$, $A_0\ne0$), the regular band structure transforms: the trivial gaps close and non-trivial gaps reopen. These states are topologically protected and no edge-defect, disorder or corner propagation can backscatter these states due to the lack of counter edge state given the spin. Notice that there are two different edge states with distinct group velocities in one gap. In contrast to honeycomb lattice which has two arm-chair edges, dimertized Lieb lattice can support two distinctive types of edges in one semi-infinite strip. Importantly, compared with standard lieb lattice($\kappa_1= \kappa_2$, $A_0\ne0$), the central flat band of dimertized Lieb lattice becomes perturted and dispersive due to the lower local symmetry. 

\begin{figure}[htbp]
\centering
\includegraphics[width=\linewidth]{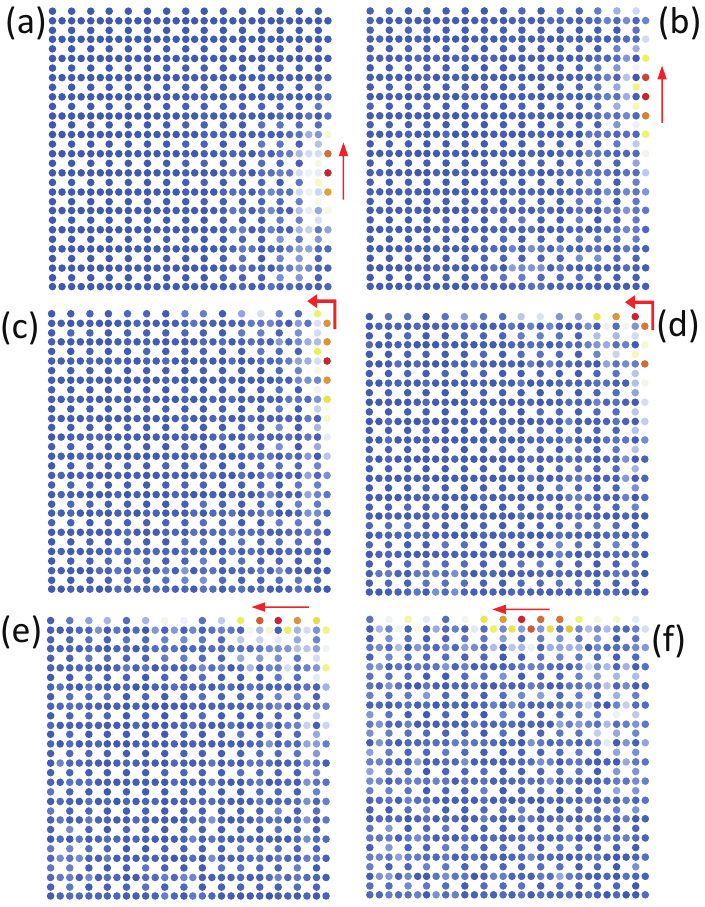}
\caption{Unidirectional edge states. Topological protected edge states can pass around the corner without any reflection. The figures (a-f) are drawn in order of propagation steps along $z$.}
\label{fig:unidirection}
\end{figure}

Tight-binding calculations of dimertized Lieb photonic topological insulator expose the topological protection of its edge states. We excite one edge state at edge of square Lieb lattices. The numerical results in Fig.(\ref{fig:unidirection},\ref{fig:defects}) show that the edge state propagate unidirectionally and immune with the corners and edge defects. 

\begin{figure*}
\centering
\includegraphics[width=\linewidth]{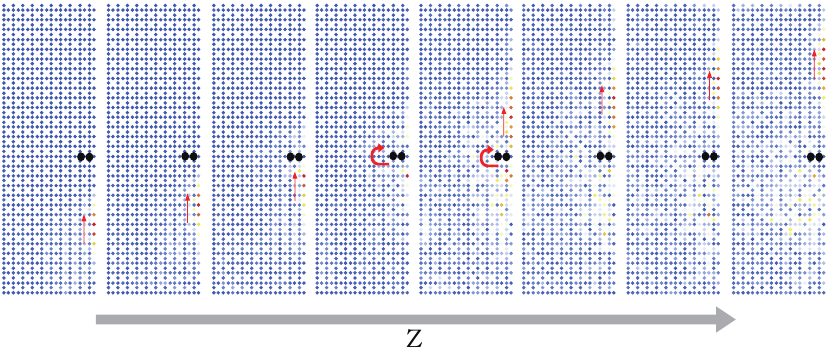}
\caption{Robustness of edge states. Black points represent the edge defects. Topological protected edge states can pass around the edge defect. The gray arrow indicates the propagation evolutions of figures along $z$.}
\label{fig:defects}
\end{figure*}

\end{widetext}

\end{document}